\documentclass{article}

\usepackage[utf8]{inputenc}
\usepackage[margin = 1in]{geometry}
\usepackage{amsmath}
\usepackage{amsfonts}
\usepackage{amssymb}
\usepackage{amsthm}
\usepackage{mathtools}
\usepackage{physics}
\usepackage{overpic}
\usepackage{epstopdf}
\usepackage{graphicx}
\usepackage[hidelinks]{hyperref}
\usepackage{bm}
\usepackage[font = small]{caption}
\usepackage[font = small]{subcaption}

\usepackage{ulem}

\graphicspath{{figures/}{../figures/}}

\newcommand{\beq}{\begin{equation}\begin{aligned}}
\newcommand{\eeq}{\end{aligned}\end{equation}}

\newcommand{\eps}{\ensuremath{\varepsilon}}
\newcommand{\Ord}[1]{\ensuremath{\mathcal{O}\left( #1\right)}}

\renewcommand \O{\mathcal{O}}

\usepackage[colorinlistoftodos,textsize=tiny,obeyFinal]{todonotes}

\allowdisplaybreaks

\newcommand \bl{\color{black}}%{\color{blue}}

\newcommand \rd{\color{black}}%{\color{red}}
\newcommand \ii{\mathrm{i}}

\title{Integral Asymptotics, Coalescing Saddles, and Multiple-scales Analysis of a Generalised Swift-Hohenberg Equation}
\author{V\'{a}clav Klika\footnote{Department of Mathematics, FNSPE, Czech Technical University in Prague, Trojanova 13, 120 00 Praha, Czech Republic}, Mohit P.~Dalwadi\footnote{Mathematical Institute, University of Oxford, Oxford OX2 6GG, UK}, Andrew L.~Krause\footnote{Mathematical Sciences Department, Durham University, Upper Mountjoy Campus, Stockton Rd, Durham DH1 3LE, United Kingdom}, Eamonn A.~Gaffney$^\dagger$.}
\date{\today}

\begin{document}
%\listoftodos[ToDos]
%\newpage

\maketitle

\begin{abstract}
% \todo{please add your names and institutions}
 Integral asymptotics play an important role in the analysis of differential equations and in a variety of other settings. In this work, we apply an integral asymptotics approach to study spatially localized solutions of a heterogeneous generalised Swift-Hohenberg equation. The outer solution is obtained via WKBJ asymptotics, while the inner solution requires the method of coalescing saddles. We modify the classic method of Chester et al.~to account for additional technicalities, such as complex branch selection and local transformation to a cubic polynomial. By integrating our results, we construct an approximate global solution to the generalised Swift-Hohenberg problem and validate it against numerical contour integral solutions. We also demonstrate an alternative approach that circumvents the complexity of integral asymptotics by analyzing the original differential equation directly through a multiple-scales analysis and show that this generates the same leading-order inner solution obtained using the coalescing saddles method  {\bl at least for one of the cases considered via integral asymptotics}. %This suggests that, for some classes of problems, the multiple-scales approach may serve as a simpler alternative.
{\bl Our findings reinforce the significance of integral asymptotics in approximating the fourth order  differential equations found in the linear stability analysis for generalisations of the Swift Hohenberg equations. This study has  also highlighted a conjecture  that, in certain cases, the method of coalescing saddles can be systematically replaced by multiple-scales analysis using an intermediary differential equation, a hypothesis for future investigation.}

\end{abstract}

%%%%%%%%%%%%%%%%%%%%%%%%%%%%%%%%%%%%%%%%%%%%%%%%%%%%%%%%%%%%%%%%%%%%%%%%%%%%%%%%
%%%%%%%%%%%%%%%%%%%

\section{Introduction}

Integral asymptotics are of use not only in the approximation of integrals themselves \cite{olver1997asymptotics,wong2001asymptotic,temme2014asymptotic} but also in determining the asymptotic behaviour of solutions to linear differential equations \cite{Bender1999,white2005asymptotic,temme2014asymptotic,wong2001asymptotic}. In this article, we consider  the latter {\bl scenario} and analyse the inner region behaviour {\bl for the linear instability problem associated with a generalisation of the Swift-Hohenberg equation that is also heterogeneous,  where the outer solution is calculated} using the WKBJ method. To resolve the inner solution fully, we employ the method of coalescing saddles \cite{chester1957extension}, which requires a detailed analysis. This includes the selection of the appropriate complex branch in the process of identifying a suitable local transformation to the cubic polynomial, which is central to the method of Chester et al. \cite{chester1957extension}. These techniques have been utilised and adapted to investigate diffraction catastrophes in optics \cite{ludwig1966uniform,berry1980iv}, as well as in more general asymptotic theory involving the higher-order Stokes phenomenon \cite{howls2004higher}.

We additionally demonstrate that, in {\bl a  particular instance, it is possible to  reproduce the results of  the complex integral asymptotics of coalescing saddles by  reverting} to the original differential equation, and then {\bl employing a suitable scaling with the method of multiple scales}  \cite{Bender1999,kuehn2015multiple,white2005asymptotic} to obtain the inner solution. We hypothesise that this approach may be applicable to a broader class of problems, as a simpler alternative to the coalescing saddles method.

\subsection{Swift-Hohenberg as a model of pattern formation}

Spatial and spatiotemporal patterns in nature occur across biological and physical media, spanning numerous temporal and spatial scales of organisation. Examples occur in fluid dynamics, solidification fronts, nonlinear optics, and biological media, which are studied from a range of perspectives \cite{cross1993pattern, ball1999self}. While the idea of understanding such phenomena using the notion of an `instability' goes back to work in fluid dynamics in the 19th and early 20th century \cite{gallaire2017fluid, drazin2002introduction}, and similar ideas in biological settings date to Turing's work in the 1950s \cite{turing1952chemical}, it is really in the last half century or so that a modern blend of numerical methods and dynamical systems theory has elucidated the importance of bifurcation points as organising centres for spatial and spatiotemporal pattern formation. A key development was the introduction of a relatively simple model for Rayleigh-Ben\'{a}rd convection, the Swift-Hohenberg equation (see \cite{swift2008swift} for an overview of the history and references to the additional literature). Recently, this relatively simple caricature model has been explored in the context of spatially varying or heterogeneous parameters. This has included a particular focus on understanding how parameter heterogeneity influences both linear stability theory and the resulting bifurcating patterns that emerge from this instability \cite{krause2024pattern}, which gives rise to localised patterns through a mechanism distinct from homoclinic snaking, {\bl with the latter  also best investigated} in the context of the Swift-Hohenberg model \cite{woods1999, burke2006, burke2009swift, bramburger2024localized}. While spatially heterogeneous states are germane to certain types of patterning instabilities in fluid dynamics \cite{harris2000inhomogeneous, soward2016frequency}, the work in \cite{krause2024pattern} originates instead from exploring biological pattern-forming systems, {\bl consisting of two coupled reaction-diffusion equations} with background spatial heterogeneity \cite{krause2020, gaffney2023, dalwadi2023universal}. The first goal of the present study is to generalise the asymptotic analysis  of the linear instability of these heterogeneous reaction diffusion models to the case of an arbitrary fourth-order linear operator, while remaining within a class of variational systems that exhibit pattern formation. This linear analysis is the first step in developing a picture of how these complicated dynamical systems behave. {\bl In pursuing this goal we also observe that that it can be possible to relate the  complicated integral asymptotics to the multiple scale analysis of differential equations, suggesting the prospect of an alternative to the method of coalescing saddles in integral asymptotics, at least in certain cases.} 

 To set the scene, we recall the classical version of the  Swift-Hohenberg equation for a field $u(t,x)$ given by
\beq\label{conservative_SH_eq}
    \partial_t u = r u - \left(1 + \varepsilon^2 \frac{\partial^2}{\partial x^2}\right)^2 u - N(u), \quad x \in [0,L],
\eeq
where $r$ is a linear bifurcation parameter controlling instability, the parameter $0<\varepsilon \ll 1$ is related to a separation of scales between the high spatial  frequency of emergent patterns compared to any heterogeneous forcing, and is henceforth considered asymptotically small. In addition,  $N(u)$ is a nonlinear term needed to dissipate growing instabilities; we will henceforth assume that $N(0)=N'(0)=0$ so that $u=0$ is the main equilibrium of interest. Throughout the article, we assume that $u$ satisfies the generalised Neumann boundary conditions of the form,
\beq\label{BCs_eq}
    \frac{\partial u}{\partial x}(t,0) = \frac{\partial u}{\partial x}(t,L) = \frac{\partial^3 u}{\partial x^3}(t,0) = \frac{\partial^3 u}{\partial x^3}(t,L) =0.
\eeq
Hence, linearising \eqref{conservative_SH_eq} around $u=0$ and using the ansatz $u \propto e^{\lambda t + \ii kx}$ we have that  the dominant spatial modes possess a spatial frequency of {\bl $  k_c = \varepsilon^{-1}$ }  and  grow at a rate $\lambda \sim r$, and so we expect these to lead to nontrivial spatial solutions for $r>0$. As Equation \eqref{conservative_SH_eq} admits a Lyapunov functional, and hence cannot exhibit temporally complex asymptotic behaviour, such instabilities typically lead to stationary spatial patterns, due either to supercritical or subcritical instabilities (depending on the form of $N(u)$) at the bifurcation point $r=0$. The authors in \cite{krause2024pattern} considered the case of a spatially varying parameter $r(x)$, and exploited the asymptotic parameter $\varepsilon \ll 1 $ to show that linear instabilities gave rise to localised patterns approximately confined to regions where $r(x)>0$, as long as the nonlinearity locally (i.e.~naively treating the spatial variable as a parameter) corresponds to a supercritical bifurcation at $r=0$. This confinement is asymptotic in $\varepsilon$, and for any finite value one observes a boundary layer around the bifurcation point $x^*$ where $r(x^*)=0$ which was resolved via a careful asymptotic expansion.

Here we extend these ideas to a fourth-order equation which can be seen as a direct generalisation of the Swift-Hohenberg equation \eqref{conservative_SH_eq}. This takes the form,
\beq\label{main_eq}
        \partial_t u = r(x) u - \left(1 + \alpha\varepsilon^2 \frac{\partial^2}{\partial x^2}+ \beta\varepsilon^4 \frac{\partial^4}{\partial x^4}\right) u - N(u),\quad x \in [0,L],
\eeq
where {\bl  $\alpha, \, \beta \sim \mbox{ord}(1)$}  are positive constants, and again we assume that $u$ satisfies the boundary conditions given by \eqref{BCs_eq} {\bl and  that $N(0)=N'(0)=0$.}  Briefly recalling the  results of {\bl a linear stability analysis for the homogeneous parameter  case of $r(x)\rightarrow r,$ constant, we remark that dominant} spatial modes possess  frequency $k_c \sim \sqrt{\alpha(2\beta)^{-1}}\varepsilon^{-1}$ and grow at a rate $\lambda \sim r -1 +\alpha^2(4\beta)^{-1}$. We also remark that {\bl  Eqn.\eqref{main_eq} is associated with a} Lyapunov functional of the form,
\beq\label{energy}
    E(u) =  \int_0^1 \frac{1}{2}\left((1-r(x)) u^2-\alpha\eps^2\left |\frac{\partial u}{\partial x}\right |^2+\beta\eps^4\left |\frac{\partial^2 u}{\partial x^2}\right |^2\right)-\left(\int_0^u N(v)\mathrm{d}v\right)\mathrm{d}x.
\eeq
Hence, while this model still admits a variational structure, and so does not exhibit temporal variation for large time, it differs from the classical Swift-Hohenberg equation in that $\alpha$ and $\beta$ independently varying now allow for control over different mechanisms of instability and dissipation.  In particular, $\alpha$ captures a short-range instability, and $\beta$ a long-range stabilising (hyperdiffusive) effect, {\bl reminiscent of the interaction between short-range activation and long-range inhibition that characterises the Turing instability.}  %\sout{\ak{Can a sentence be added to differentiate the asymptotic work needed to do this generalisation from that in the previous paper? Is there a technical obstruction here which makes it distinct in an interesting way?}} 
In addition, we analyse both turning points, showing quite a distinct behaviour, and instead of numerically choosing the correct branch {\bl from the possible choices when developing the integral asymptotics,} here we provide detailed arguments leading to the branch selection analytically.

The goal of this paper is twofold: (i) to construct asymptotic solutions of the linearised form of \eqref{main_eq}, and in particular resolve the behaviour across turning points using the coalescing saddles method which demarcate regions of localised patterning; (ii) propose a conjecture that the coalescing saddles method can be circumvented by a multiple-scales approach, which we illustrate on the very same problem. To this end, in Section \ref{sec_asymptotics} we construct the outer solution of the linearised problem using a WKBJ ansatz, and begin developing an inner solution for use at turning points by writing the problem in terms of a contour integral formulation. In Section \ref{sec_turning_pts} we focus on resolving the behaviour across turning points using the method of coalescing saddles, which we then compare in Section \ref{sec_numerical} to a numerical evaluation of the contour integral. We give an alternative way to arrive at the solution across the turning points using a more direct multiple-scales approach in Section \ref{sec.multiple_scales}. Finally, we close in Section \ref{discussion} by discussing related open problems.

\section{Asymptotics of generalised pattern forming model}\label{sec_asymptotics}
    % Consider the following non-conservative variant of Swift-Hohenberg model as discussed above:
    % \begin{align*}
    %     \partial_t u = r(x) u - \left(1 + \alpha\varepsilon^2 \frac{\partial^2}{\partial x^2}+ \beta\varepsilon^4 \frac{\partial^4}{\partial x^4}\right) u - N(u, x),\quad \alpha>0, \beta>0,
    % \end{align*}
    % where we used a suitable normalisation in the second linear term on the right while the localisation effect is restricted to the linear term with no derivatives. Note that the typically considered conservative case, when the governing equation is of a gradient form with conserved ``energy'', corresponds to the special case when $\alpha=2\beta$ (and rescaling the spatial coordinate by $\sqrt{\alpha}$). In addition, the parameter $\varepsilon$ is considered small below and is an effective measure of the shallowness of $r(x)$ slope, $N(u,x)$ stands for a nonlinearity.

    % The Swift-Hohenberg model is known to be a generic pattern formation model including a localised pattern arising from a subcritical homoclinic snaking even for the conservative case with $r$ constant. Here, we shall analyse instability due to the spatial heterogeneity in the linear kinetic term $r$ in the supercritical case relaxing the condition of conservability.

    Focusing on the linearised problem (i.e.~dropping $N(u)$), and assuming a separation of variables in time and space, $u(t,x)=e^{\lambda t} p(x)$, we find that $p(x)$ satisfies
    \begin{equation} \label{eq.p}
      0=(r(x)-\lambda)p - \left(1+\alpha\varepsilon^2 \frac{\dd^2}{\dd x^2}+\beta\varepsilon^4 \frac{\dd^4}{\dd x^4}\right)p.
    \end{equation}
We will chiefly be interested in cases where $\lambda>0$, corresponding to instability of the full problem. We assume throughout that $r(x)$ is sufficiently smooth and does not vary too rapidly, i.e.~we assume that $|r'(x)| = O(1)$ for all $x\in [0,L]$.

\subsection{Outer solution}
    The outer solution, away from turning points, can be readily found using the WKBJ ansatz, $p(x)=e^{\frac{i}{\varepsilon}h(x)}(S_{0}(x)+\varepsilon S_{1}(x)+\ldots)$, to arrive at the leading-order solution    
   {\bl
    \begin{eqnarray}\label{eq.p_out}
        p_{\rm{out}}(x) &\sim& \frac{1}{\left|h_+'(x)(\alpha-2 \beta h_+'^2(x))\right|^{1/2}}\left(A_+ \cos\left(\frac{1}{\varepsilon}h_+(x)\right)+B_+ \sin\left(\frac{1}{\varepsilon}h_+(x)\right)\right) \\ \nonumber &+& \frac{1}{\left|h_-'(x)(\alpha-2 \beta h_-'^2(x))\right|^{1/2}}\left(A_- \cos\left(\frac{1}{\varepsilon}h_-(x)\right)+B_- \sin\left(\frac{1}{\varepsilon}h_-(x)\right)\right),.
    \end{eqnarray}
   Here, 
    \begin{equation*}
      h_\pm(x) =  \left(\frac{\alpha}{2 \beta}\right)^{1/2} \int_y^x \left(1\pm \left[\left(1-4\frac{\beta}{\alpha^2}\right)+4 \frac{\beta}{\alpha^2}(r(z)-\lambda)\right]^{1/2}\right)^{1/2}\dd z,
    \end{equation*}
   and   $A_\pm,~B_\pm$ are constants determined via imposed boundary conditions, with $y$ also a constant that can be set to any   convenient value without loss, noting that a change in $y$ can be accommodated by a change in $A_\pm,B_\pm.$}  However, the outer solution blows up at the so-called ``turning points''
    \begin{align*}
        x = a &: \quad r(a) - \lambda = 1-\frac{\alpha^2}{4 \beta}<1 \mbox{ corresponding to } \alpha-2\beta h_\pm'(a)^2=0,
        \\
        x = b &: \quad r(b) - \lambda = 1 \mbox{ corresponding to } h_\pm'(b)^2=0.
    \end{align*}
  Hence, the turning point $x=a$ corresponds to the lower value of $r$ at the two turning points.

  % One can show that the regions of validity of the outer solution is $|x-a|\gg $\todo{vk: add once appendix realculated}, see Appendix \ref{app.RegionsWKBJvalidity}.
  Note that the usual stability requirement of homogeneous stability of the steady state in Turing's theory requires $r(x)-1=\lambda<0$, so that the second turning point is typically neglected in studying pattern-forming instabilities (as was done in \cite{krause2024pattern}). Here we will also analyze such turning points in order to fully understand the structure of global solutions.

 To be able to match boundary conditions irrespective of their location with respect to the turning points, we need to identify the solution's behaviour across the turning points. For this purpose, we shall use asymptotic methods of integrals. 
 
\subsection{Inner solution fundamentals}

\subsubsection{Turning point $x=a$.}
We rewrite the differential equation \eqref{eq.p} in integral form close to the turning point $x=a$. To this end, we locally expand the spatial heterogeneity
\begin{equation*}
  r(x) = r(a) + \rho(x-a)+\Ord{(x-a)^2},\quad r(a)=\lambda+1-\frac{\alpha^2}{4 \beta},\quad \rho=r'(a).
\end{equation*}
%and rescaling $y=(x-a)/\varepsilon$, the solution $p$ satisfies (near the turning point)
%\begin{equation} \label{eq.inner}
%  0= \left(1-\frac{\alpha^2}{4\beta}\right)p + \varepsilon y \rho p - \left(1+\alpha \frac{\dd^2}{\dd y^2}+\beta \frac{\dd^4}{\dd y^4}\right)p.
%\end{equation}
We extend the governing equation to the complex plane by considering the generalisation to $x \in \mathbb{C}$ and $p(x):\mathbb{C}\to\mathbb{C}$ denoting a holomorphic function. We also rescale $z=(x-a)/\varepsilon$, so near the turning point at $x = a$ the solution $p$ satisfies
\begin{equation} \label{eq.inner}
  0= \left(1-\frac{\alpha^2}{4\beta}\right)p + \varepsilon z \rho p - \left(1+\alpha \frac{\dd^2}{\dd z^2}+\beta \frac{\dd^4}{\dd z^4}\right)p.
\end{equation}
We note that both the real and imaginary parts of this complex $p$ generate solutions to the inner problem of Eq. \eqref{eq.p}. In particular, the real and imaginary parts of a complex solution to Eq. \eqref{eq.inner},  evaluated on the real numbers will generate two linearly independent solutions, $\Re(p(\Re(z)))$ and $\Im(p(\Re(z)))$, to the original problem in Eqn. \eqref{eq.p}. Next, we consider an integral form of the solution due to the linearity of the problem
\begin{equation*}
  p(z) = \int_C f(t) e^{-zt} \dd t,
\end{equation*}
where $t\in\mathbb{C}$ and the contour $C$ is to be identified. Then,
\begin{equation*}
  zp(z) = \int_C f(t) \left(-\frac{d}{dt} e^{-zt}\right) \dd t = \int_C \frac{d f}{dt} e^{-zt} \dd t +\left[-f(t) e^{-zt}\right]^{C_{\rm end}}_{C_{\rm start}}
\end{equation*}
and hence \eqref{eq.inner} becomes
%\ak{Worth checking the next equation below carefully. Namely, there is an abuse of notation between $y$ and $z$ as the original equation has a term of the form $yp(y)$ which would be $zp(z-is)-isp(z-is)$. Shifting the argument is fine (just a choice of contour) but unclear the second term in the equation goes away.}
\begin{equation*}
  0=\eps \rho  \left[-f(t) e^{-zt}\right]^{C_{\rm end}}_{C_{\rm start}}+\int_C  e^{-zt} \left(\eps \rho \frac{d f}{dt} - (\frac{\alpha^2}{4\beta}+\alpha t^2 + \beta t^4)f \right)\dd t.
\end{equation*}
Setting the integrand to $0$ and solving for $f(t)$, we find that $p$ is a solution to the complex extension of the inner problem, Eq.~\eqref{eq.inner}, if
\begin{align}
  p(z) &=\int_C \exp\left[\frac{1}{\varepsilon \rho} \left( \frac{\alpha^2}{4 \beta}t + \frac{\alpha}{3} t^3+\frac{\beta}{5}t^5\right)-zt \right] \dd t,\nonumber\\
  0 &= \exp\left[\frac{1}{\varepsilon \rho} \left( \frac{\alpha^2}{4 \beta}t + \frac{\alpha}{3} t^3+\frac{\beta}{5}t^5\right)-zt \right]^{C_{\mathrm{end}}}_{C_{\mathrm{start}}},\label{eq.pVanishingCond}
\end{align}
where we have set the constant of integration pre-multiplying the integral to one for simplicity. To force the boundary term to vanish, we choose the contour $C$ so that it starts and ends at limits of $t$ where 
$$ \Re \left[ \frac{1}{\eps\rho}\left(\frac{\alpha^2}{4\beta} t+\frac{\alpha}{3} t^3 + \frac{\beta}{5} t^5\right) -zt \right] \to 0  $$
for all real $z$. We note that $C$ cannot be closed if we wish to obtain a nontrivial solution for $p$ (which follows from Cauchy's integral theorem). In summary, if $p(z)$ solves the complex version of \eqref{eq.inner}, then both $\Re(p(\Re(z)))$ and $\Im(p(\Re(z)))$ solve the original (real) problem \eqref{eq.p}. Hence, if we can find two independent solutions of the complex governing equation \eqref{eq.inner}, we will generate all four linearly independent solutions of the original real problem \eqref{eq.p}.

In principle, we now have solutions to the original problem \eqref{eq.inner} in terms of the contour integral, which can be used to construct the inner solution of the leading-order WKBJ asymptotics via the linearly independent solutions. However, we would like an explicit form of these solutions, at least as an approximation of the contour integral representations, to construct the inner solution and match it with the outer solutions. We now proceed using the steepest descent method (an extension of the Laplace method of integral asymptotics) for large values of $|z|$ in order to simplify this representation to a form suitable for interpretation and matching to the outer solution. %We divide this into $\Lambda=\varepsilon \rho z>0$ and $\Lambda<0$, as these will correspond to being on different sides of the turning point.

To this end, we rewrite the integral form of $p(z)$ with $\varepsilon^{*}=\varepsilon \rho$ and using the substitution $t= \ii s$ %and $\Lambda(z)=\rho\varepsilon z$  
as
\begin{equation} \label{eq.innerp}
  p(z) = \ii \int_{\tilde{C}} \exp \left(\frac{1}{\epsilon^*} \psi(s; z)\right) \dd s,\quad \psi(s;z) = \ii \left[ \left(\frac{\alpha^2}{4 \beta}-\varepsilon^* z\right)s - \frac{\alpha}{3} s^3+\frac{\beta}{5}s^5\right],
\end{equation}
and calculate the imaginary and real parts of $\psi$ for $z\in\mathbb{R}$
\begin{align} 
  \Im \psi &= s_r \left[\left(\frac{\alpha^2}{4 \beta}-\varepsilon^* z\right)-\frac{\alpha}{3}(s_r^2-3s_i^2)+\frac{\beta}{5}(s_r^4-10s_r^2s_i^2+5 s_i^4)\right],\label{eq.ReImPsi-im}\\
  \Re \psi &= -s_i \left[\left(\frac{\alpha^2}{4 \beta}-\varepsilon^* z\right)-\frac{\alpha}{3}(3s_r^2-s_i^2)+\frac{\beta}{5}(5s_r^4-10s_r^2s_i^2+ s_i^4)\right],\label{eq.ReImPsi-re}
\end{align}
where we integrate over the contour given by $s=s_{r}+i s_{i}$ with $s_{r},\,s_{i}\in\mathbb{R}$.

Next, for the leading-order approximation, we need to determine the saddles together with the direction of the steepest descent contour (SDC) at the saddles and the asymptotes away from saddles. The four saddles are
\begin{equation*}%\label{eq.saddles_x=a}
 s=\pm\left(\frac{\alpha}{2\beta}\right)^{1/2}\left(1\pm 2\kappa\right)^{1/2},
\end{equation*}
%\begin{align}\label{eq.saddles_x=a}
  %s_{++}&=\left(\frac{\alpha}{2\beta}\right)^{1/2}\left(1+2\kappa\right)^{1/2},\quad s_{-+}=-s_{++},\\
  %s_{+-}&=\left(\frac{\alpha}{2\beta}\right)^{1/2}\left(1-2\kappa\right)^{1/2},\quad s_{--}=-s_{+-},
%\end{align}
where we denote $\kappa = \sqrt{\beta\varepsilon^* z}/\alpha$. We denote two of these saddles  as 
\begin{equation} \label{eq.saddles_x=a}
 s_{\pm}=\left(\frac{\alpha}{2\beta}\right)^{1/2}\left(1\pm 2\kappa\right)^{1/2}.
\end{equation}
Recalling that we are interested in the inner solution, %where $|z|\ll1$ (more preciselly, $|\varepsilon^* z|\ll 1$), for $z>0$ small both $s_{+},~s_{-}$ are real, while for $z<0$ small both saddles are complex.
for $z>0$ both $s_{+},~s_{-}$ are real, while for $z<0$ both of these saddle locations are complex.

The SDC can be identified by tracing a curve in the complex plane where $\Im\psi$ has the same value as at the considered saddle. With the simple polynomial form of $\psi$, such a curve can be readily identified and parametrised \bl{(corresponding to a curve $s_r,~s_i$ in the complex plane given by Eq.~\eqref{eq.ReImPsi-im} with $\Im\psi(s)= \psi(s_+) $ and $ \Im\psi(s)=\Im\psi(s_-)$ for the respective saddles)}. However, this explicit knowledge is not necessary to identify the asymptotic approximation, since the leading-order contribution arises from the tangent to the SDC at the saddle, which can be determined from the second derivative of $\psi$ at the saddle. Specifically, the tangent is equal to the value of $\mathrm{arg}\left[(-\psi'')^{-1/2}\right]$ at the saddle. Near $x=a$ we have, for $z>0$ (real saddles),
\begin{equation*}
  (\psi''(s_{+}))^{-1/2}\propto (-i)^{-1/2},
\end{equation*}
where the constant of proportionality is real, and hence the angle of the SDC at $s_{+}$ is $\mu_{+}=\pi/4$. Similarly, the tangent of the SDC at $s_{-}$ is $\mu_{-}=3\pi/4$. For $z<0$ (complex saddles), the angle is nonconstant and the general expression $\mathrm{arg}\left[(-\psi'')^{-1/2}\right]$ cannot be substantially simplified, as we will describe below.

Finally, in order to satisfy the vanishing contribution from the end points of the contour, Eq.~\eqref{eq.pVanishingCond}, the solution must vanish for large enough $z$. As $\psi(s;z)\sim \ii s^{5} / 5$ for large $s$, this condition translates into the requirement that the integration contour has asymptotes whose tangents are the fifth roots of imaginary unity. In particular, for a convergent integral as $s\to+\infty$ i.e.~$\Re(\psi(s))\to -\infty$, we require
\begin{equation*}
  \exp \left( \dfrac{\psi(s)}{\eps^*} \right) \sim \exp\left(\frac{i}{5 \eps^*} |s|^5 e^{5 \ii  \mathrm{arg}(s)}\right) \to 0 \mbox{ as }|s|\to+\infty.
\end{equation*}
Hence, we require $e^{5 \ii \mathrm{arg}(s)}= \ii$ and thus $\mathrm{arg}(s)= \pi/10 + 2 \pi k/5$ for $k\in\{0,1,\ldots,4\}$ for large $|s|$.

\begin{figure}
\centering
\begin{subfigure}{.485\textwidth}
   \centering
   \includegraphics[width=.4\linewidth]{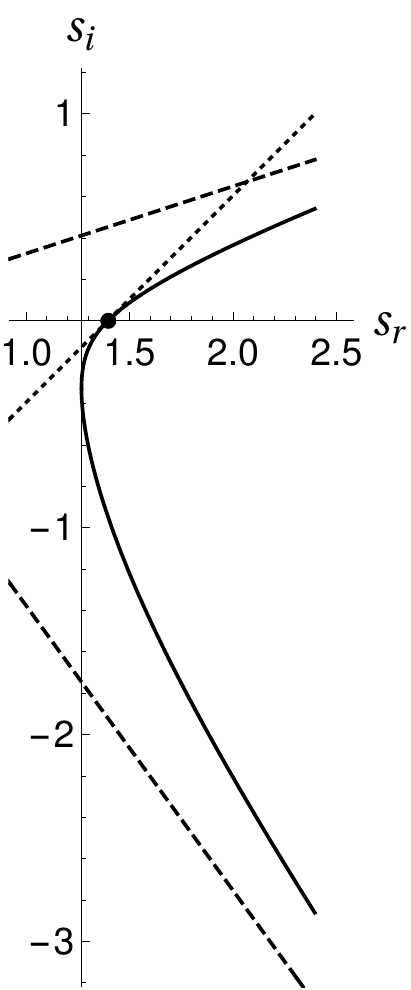}
   \caption{Plot of the steepest descent curve (with a constant phase), in the complex plane.}
\end{subfigure}~~~~
\begin{subfigure}{.485\textwidth}
   \centering
   \includegraphics[width=.95\linewidth]{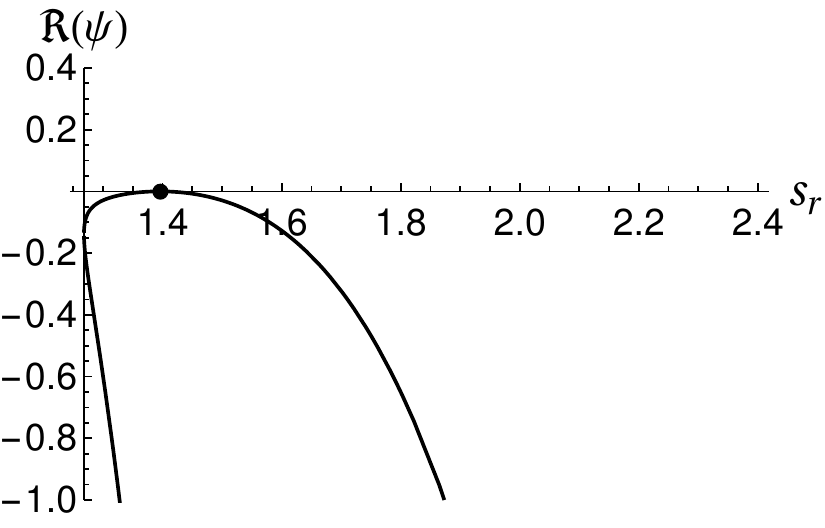}  
\caption{Real part of $\psi$ along the contour with the horizontal axis corresponds to $s_{r}$, which parameterises the curve confirming that it is the (steepest) descent curve.
}
\end{subfigure}
    \caption{\label{fig.SDC+} {\bl In plot (a) the steepest descent contour passing through $s_{+}$ is the solid curve} for $\varepsilon^* z=0.2$, $\alpha = 3$, $\beta=1$. The dashed lines indicate the revealed asymptotes and tangent at the saddle. In particular, we have $s_{+}= 1.395$, the tangent angle of the SDC at $s_{+}$ is $\pi/4$ and the asymptotes are {at the angles} $\pi/10$ and $17\pi/10$. {\bl In plot (b) the real part of $\psi$ on the steepest descent contour is given.}}
\end{figure}

\begin{figure}
\centering
\begin{subfigure}{.485\textwidth}
   \centering
   \includegraphics[width=.45\linewidth]{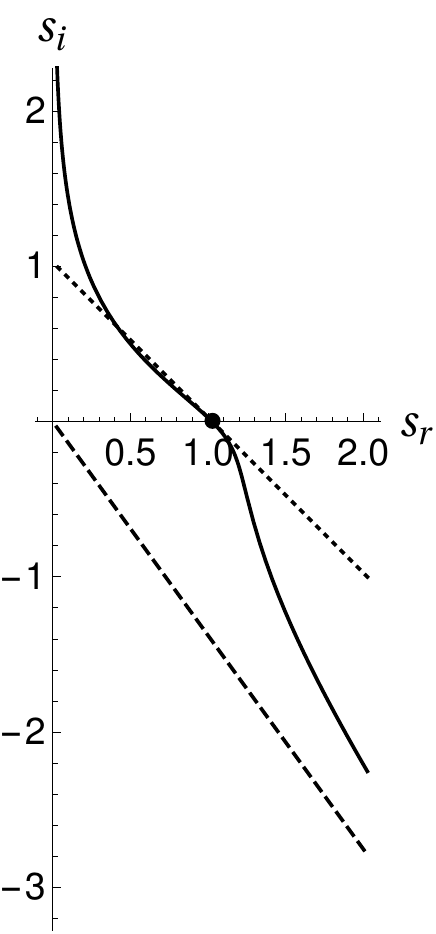}
   \caption{Plot of the steepest descent curve (with a constant phase), in the complex plane.}
\end{subfigure}~~~~
\begin{subfigure}{.485\textwidth}
   \centering
   \includegraphics[width=.95\linewidth]{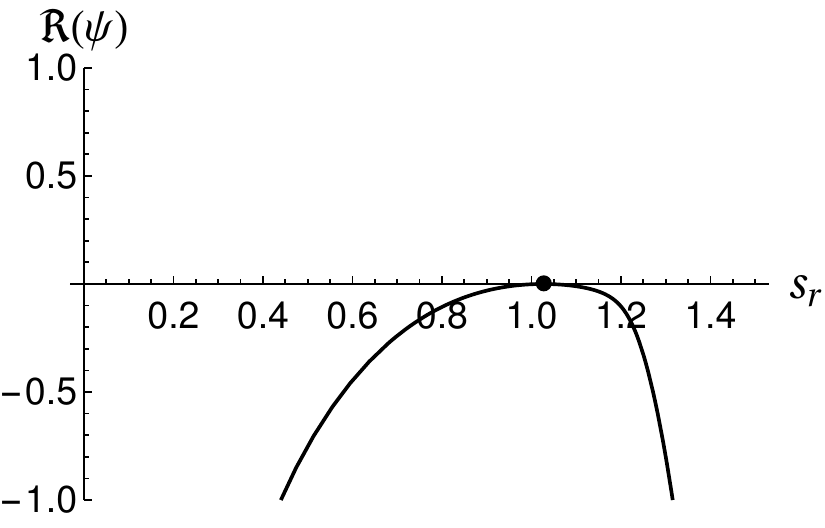}  
\caption{Real part of $\psi$ along the contour with the horizontal axis corresponds to $s_{r}$, which parameterises the curve confirming that it is the (steepest) descent curve.%{\mg VK: To be precise, this fig shows it is a descent curve only. However, if we also inspect Fig 11 in Appendix and from what we know from theory, it should be clear that it is the steepest DC.}
}
\end{subfigure}
    \caption{\label{fig.SDC-} {\bl In the plot (a)} the steepest descent contour for $\varepsilon^* z=0.2$, $\alpha = 3$, $\beta=1$ passing through $s_{-}$. The dashed lines indicate {\bl (i) the  asymptote for large $s_r$ and (ii) the tangent} at the saddle. In particular, we have $s_{-}= 1.026$, the tangent angle of the SDC at $s_{-}$ is $-\pi/4$ and {\bl the asymptote for large $s_r$ has angle $17\pi/10$, with another asymptote at angle $\pi/2.$ In plot (b) the real part of $\psi$ along the steepest descent curve is given.}}
\end{figure}

In summary, since we only require the asymptotic behaviour of the integral in Eq.~\eqref{eq.innerp}, we can use two helpful approximations. First, the SDC can be replaced by a straight line passing through the saddle in the direction of the SDC tangent; second, the argument of the exponential can be replaced by its leading-order approximation. The latter corresponds to the quadratic approximation of the argument which, by the construction of the SDC, is real and negative. We can then apply Laplace's method to the resulting approximations. To obtain the quadratic approximation, we calculate $\psi''=i(-2\alpha s+4\beta s^{3})$ at the saddles while confirming that the integrand becomes exponentially small on the SDC for large $|s|$.
%Note that the value of $\psi''$ is purely imaginary (since the saddles are real for $z>0$), and hence we expect oscillatory solutions.

\paragraph{Real case when $z>0$.} Now we are ready to find the asymptotic behaviour for the real case when $0<\varepsilon^* z=\rho(x-a)\ll1$. Note  that the value of $\psi''$ is purely imaginary  for $z>0$, since the saddles are real for $z>0$, and hence we expect oscillatory solutions here. 
Proceeding, the approximate solutions to the original problem, Eq.~(\ref{eq.inner}), then read
\begin{align} \label{eq.inner.x=a.L>0.solu}
  p_{+}(x) &\sim \frac{\sqrt{\pi \varepsilon^*}}{2^{1/4}} \frac{\beta^{1/4}}{\alpha^{3/4}} \frac{1}{\kappa^{1/2}}\frac{1}{(1+2 \kappa)^{1/4}} \cos\left[\frac{\sqrt{2}}{15}\frac{1}{\varepsilon^*}\frac{\alpha^{5/2}}{\beta^{3/2}}(1+2\kappa)^{1/2}(1-\kappa-6\kappa^2)\right]\\
  &\bl{\sim \frac{\sqrt{\pi \varepsilon^*}}{2^{1/4}} \frac{\beta^{1/4}}{\alpha^{3/4}} \frac{1}{\kappa^{1/2}} \cos\left[\frac{\sqrt{2}}{15}\frac{1}{\varepsilon^*}\frac{\alpha^{5/2}}{\beta^{3/2}}\left(1-\frac{15}{2}\kappa^2\right)\right]}, \nonumber\\
    p_{-}(x) &\sim \frac{\sqrt{\pi \varepsilon^*}}{2^{1/4}} \frac{\beta^{1/4}}{\alpha^{3/4}} \frac{1}{\kappa^{1/2}}\frac{1}{(1-2 \kappa)^{1/4}} \cos\left[\frac{\sqrt{2}}{15}\frac{1}{\varepsilon^*}\frac{\alpha^{5/2}}{\beta^{3/2}}(1-2\kappa)^{1/2}(1+\kappa-6\kappa^2)\right]\\
    &\bl{\sim \frac{\sqrt{\pi \varepsilon^*}}{2^{1/4}} \frac{\beta^{1/4}}{\alpha^{3/4}} \frac{1}{\kappa^{1/2}} \cos\left[\frac{\sqrt{2}}{15}\frac{1}{\varepsilon^*}\frac{\alpha^{5/2}}{\beta^{3/2}}\left(1-\frac{15}{2}\kappa^2\right)\right]}, \nonumber
\end{align}
recalling that $\kappa=\sqrt{\beta\rho(x-a)}/\alpha>0$ and where the subscript corresponds to the chosen saddle approximation. Note that there is another pair of approximate solutions when replacing $\cos$ with $\sin$ in the expressions above. The former corresponds to the real part of \bl{$p(x)$} while the latter corresponds to the imaginary part. Together, this yields four independent solutions, as expected (noting that the second pair of saddles does not yield new independent solutions due to symmetry). To illustrate this, we simplify the asymptotic expression for the real part of the solution as follows:
\begin{multline*}
  \Re(p_{+}(\Re z)) \sim \Re \left[\exp\left(\frac{\psi_{+}}{\varepsilon*}\right)\int_R \exp\left(-\frac{1}{\varepsilon*}\underbrace{\left(-\frac{\psi''_{+}}{2} e^{2 \ii \mu_{+}}\right)}_{>0}r^2\right)\dd r\right]\\=\left(\int_R \exp\left(-\frac{1}{\varepsilon*}\left(-\frac{\psi''_{+}}{2} e^{2 \ii \mu_{+}}\right)r^2\right)\dd r\right) \cos\left(\frac{1}{\varepsilon*}\psi_{+}\right),
\end{multline*}
where $\mu_{+}=\pi/4$ is the tangent of the SDC at the $s_{+}$ saddle and the subscript $+$ denotes evaluation of the functions $\psi,~\psi''$ at the  $s_{+}$ saddle. 

Observe that all four solutions are singular at the turning point $x=a$, and hence we need to resolve the inner region further, noting that the identified inner solution matches the WKBJ outer solution (the former is a linearisation of the spatial dependency close to the turning point of the latter).

% imag case, Lambda<0
\paragraph{Imaginary case $z<0$.} 
Next, we focus on the case when $z<0$, which corresponds to complex saddles. The situation is similar to, but more technically involved than, the real case we discussed above. Again, the leading-order approximation arises from the SDC and is governed by the contributions from the immediate tangent neighbourhood of the saddles. Hence, we may simplify the SDC by a line with the same tangent as the SDC at the saddle with the largest real part of $\psi$. We therefore calculate the saddle positions, and then at these saddles the local tangent of the SDC and the quadratic approximation of $\psi$. Recall that $\psi$ was purely imaginary at the saddles in the case $z>0$, studied above, and hence the solution was oscillatory. In the case of $z<0$, when saddles are complex, the solution will be oscillatory with an exponential envelope.

We start by rewriting the saddles, Eq.~\eqref{eq.saddles_x=a}, so that their real and imaginary part are explicit:
\begin{align}\label{eq.saddles_x=a_imag}
  s_{+}&=\left(\frac{\alpha}{2\beta}\right)^{1/2}\left(1+4\kappa^2\right)^{1/4}(\cos\varphi+i \sin\varphi),\\
  s_{-}&=\left(\frac{\alpha}{2\beta}\right)^{1/2}\left(1+4\kappa^2\right)^{1/4}(\cos\varphi-i \sin\varphi)=\overline{s_{+}},\\
\end{align}
where $\varphi=\frac{1}{2} \mathrm{arctan}(2\kappa)$, $\kappa=\sqrt{-\beta\varepsilon^* z}/\alpha$ and again we have another pair of saddles $s_{-+}=-s_{+},~s_{--}=-s_{-}$ which we do not consider further due to the symmetry.

The SDC tangent at the saddles is simply $\mu = \mathrm{arg}\left((-\psi'')^{{-1/2}}\right)$ evaluated at the saddles, noting that its value is not constant for the $z<0$ case we consider here, contrary to the positive case. The real and imaginary parts of $\psi$ follow from eqns. (\ref{eq.ReImPsi-im}), (\ref{eq.ReImPsi-re}) and the value of $-\psi''$ at  the saddles that yield the quadratic approximation follows analogously. By direct inspection of the SDC of both the $s_{+}$ and $s_{-}$ saddles it follows that the two saddles $s_{+},~s_{-}$ lie on the same contour $\{s_{r},s_{i}\}\in\mathbb{C}$ given by Eq.~\eqref{eq.ReImPsi-im}
\begin{multline} \label{eq.Imx=aImagCase}
 s_r \left[\left(\frac{\alpha^2}{4 \beta}-\varepsilon^* z\right)-\frac{1}{3}\alpha(s_r^2-3s_i^2)+\frac{1}{5}\beta(s_r^4-10s_r^2s_i^2+5 s_i^4)\right]=\\
 =  \Im\psi_+=\Im\psi_-=\frac{\alpha^{5/2}}{15 \sqrt{2} \beta ^{3/2}} (1+4 \kappa ^2)^{1/4} \left(1+\sqrt{4 \kappa ^2+1}+12 \kappa ^2\right) \cos\varphi \bl{\sim \frac{\sqrt{2} \alpha^{5/2}}{15  \beta ^{3/2}}(1+8\kappa^2)\cos\varphi},
\end{multline}
where $\kappa=(-\beta\varepsilon^* z)^{{1/2}}/\alpha$. In addition, $\Re(\psi)$ is larger at $s_{-}$, in fact,
\begin{equation} \label{eq.Rex=aImagCase}
0<-\Re(\psi_{+})=\Re(\psi_{-})=\frac{\alpha^{5/2}}{15 \sqrt{2} \beta ^{3/2}} (1+4 \kappa ^2)^{1/4} \left(1-\sqrt{4 \kappa ^2+1}+12 \kappa ^2\right) \sin\varphi \bl{\sim \frac{\sqrt{2} \alpha^{5/2}}{3 \beta ^{3/2}}  \kappa^2 \sin\varphi}.
\end{equation}

\begin{figure}
\centering
\begin{subfigure}{.485\textwidth}
   \centering
   \includegraphics[width=.9\linewidth]{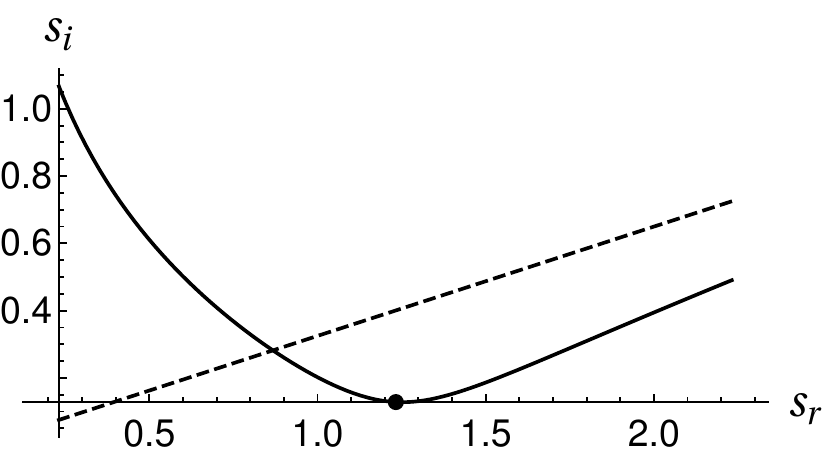}
   \caption{Plot of the steepest descent curve (with a constant phase), in the complex plane.}
\end{subfigure}~~~~
\begin{subfigure}{.485\textwidth}
   \centering
   \includegraphics[width=.95\linewidth]%{TurningPointa_LambdaNeg_s+_alpha3,beta1_re_insetAB.pdf}
{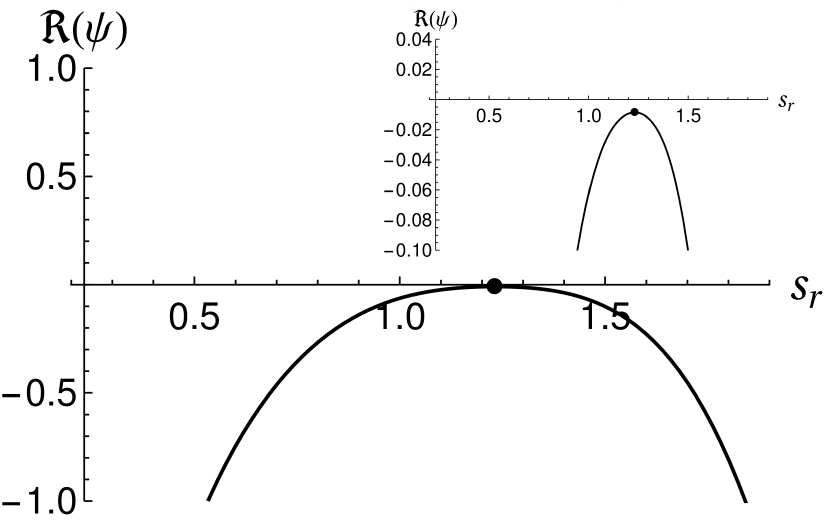}  
\caption{Real part of $\psi$ along the contour with the horizontal axis corresponds to $s_{r}$, which parameterises the curve confirming that it is the (steepest) descent curve. \bl{The left $y$-axis corresponds to the solid while the right $y$-axis to the dashed curve. The latter shows that indeed the $0<-\Re(s_+)=\Re(s_-)$ when compared in scale to Fig \ref{fig.SDC-neg}b.}%The inset shows that indeed the $0<-\Re(s_+)=\Re(s_-)$ when compared in scale to Fig \ref{fig.SDC-neg}b.}
}
\end{subfigure}
  \caption{\label{fig.SDC+neg} {\bl In plot (a) the steepest descent contour passing through $s_{+}$ is given by the solid curve  for $\varepsilon^* z=-0.2$, $\alpha = 3$, $\beta=1$ . The dashed lines indicate the revealed asymptote for large $s_r$. In particular, we have $s_{+}= 1.231+ \ii 0.128$, the asymptote for large $s_r$ has angle $\pi/10$ with the  other asymptote at angle $\pi/2$ and the 	 tangent angle of the steepest descent contour  at $s_{+}$ given by $-0.023 \pi$, which is indistinguishable from the horizontal axis at this resolution. In plot (b) the real value of $\psi$ is given for the steepest descent contours passing through $s_+$, with the two curves corresponding to two different vertical axes.}}
\end{figure}

\begin{figure}
\centering
\begin{subfigure}{.485\textwidth}
   \centering
   \includegraphics[width=.5\linewidth]{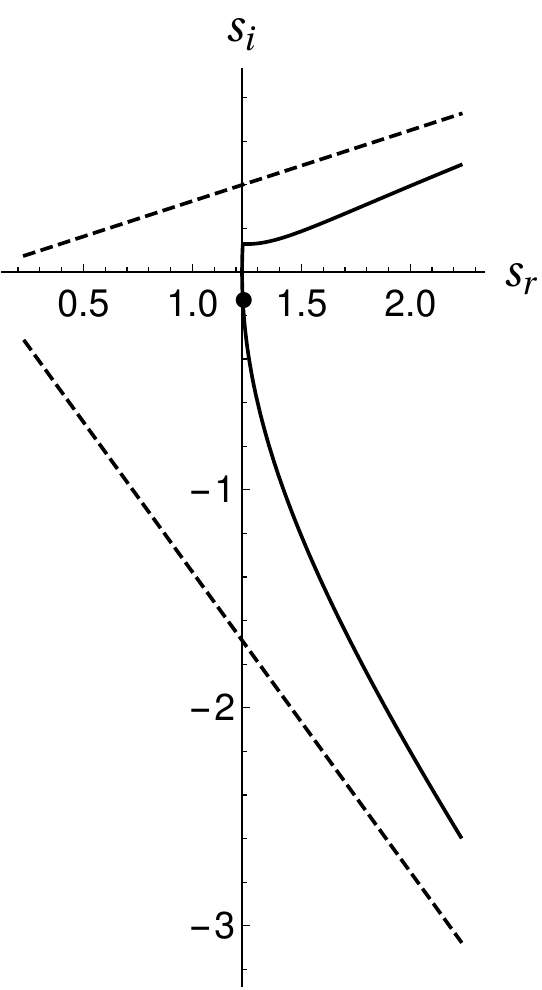}
   \caption{Plot of the steepest descent curve (with a constant phase) through $s_-$, in the complex plane. }
\end{subfigure}~~~~
\begin{subfigure}{.485\textwidth}
   \centering
   \includegraphics[width=.95\linewidth]{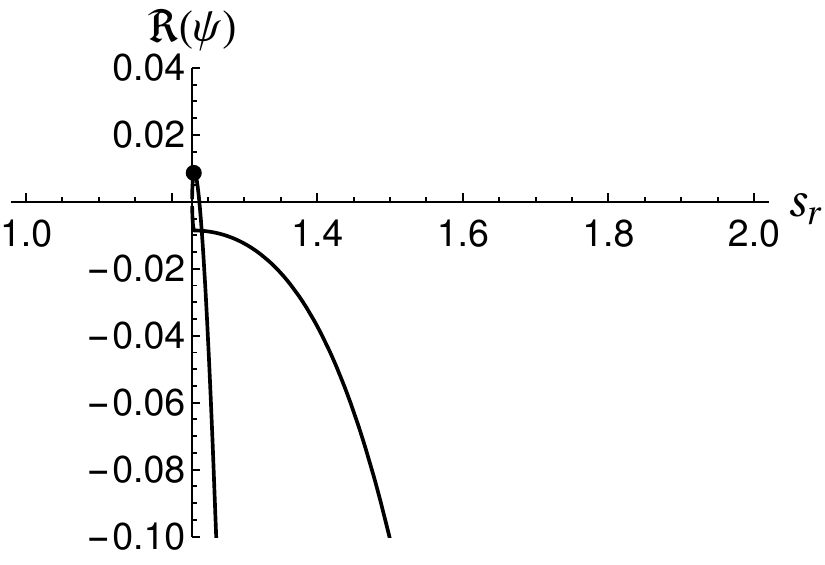}  
\caption{Real part of $\psi$ along the contour with the horizontal axis corresponds to $s_{r}$, which parameterises the curve confirming that it is the (steepest) descent curve.%{\mg VK: To be precise, this fig shows it is a descent curve only. However, if we also inspect Fig 11 in Appendix and from what we know from theory, it should be clear that it is the steepest DC.}
}
\end{subfigure}
    \caption{\label{fig.SDC-neg}   In the plot (a) the steepest descent contour for $\varepsilon^* z=-0.2$, $\alpha = 3$, $\beta=1$ passing through $s_{-}$. {\bl Note that the vertical axis is through the saddle point rather than the origin and} that the dashed lines indicate the revealed asymptotes and tangent at the saddle. In particular, we have $s_{-}= 1.231 - \ii ~ 0.128 = \bar{s}_{+}$, the tangent angle of the SDC at $s_{-}$ is {\bl $(1/2-0.023) \pi$, which is indistinguishable from the vertical axis at this resolution and the asymptotes are at the angles $\pi/10$ and $17\pi/10$.
    Also, observe  that this  steepest descent contour proceeds to 
   the saddle at $s_+=\bar{s}_-$, depicted by the dot in 
   Fig.\,\ref{fig.SDC+neg}(a). In   Fig.\,\ref{fig.SDC-neg}, there is a  discontinuity in the contour tangent at this point, and the steepest descent contours from plots Fig.\,\ref{fig.SDC+neg}(a),
   Fig.\,\ref{fig.SDC-neg}(a) subsequently coincide,   
   due to the need to select an appropriate asymptote as $|s|\rightarrow \infty.$ Finally note that the contribution of the contour integral local to the saddle $s_+$ is subordinate as $\Re(\psi(s_+))=-\Re(\psi(s_-))<0<\Re(\psi(s_-)).$ In plot (b) the real part of $\psi$ along the steepest descent contour is given.} 
    }
\end{figure}

Therefore, while the SDC through the saddle $s_{-}$ also passes through $s_{+}$,  {\bl 
as observed and detailed in Fig.~\ref{fig.SDC-neg}, the contribution of the neighbourhood of $s_{+}$ to 
the contour integral along the $s_{-}$ contour, is subordinate for the leading-order asymptotics. On the other hand, 
the SDC passing through  the saddle $s_{+}$ does not contain the saddle $s_-$ since the contour joining $s_+$ from $s_-$,  with $\Im \psi$ fixed, 
would be a steepest ascent contour and thus perpendicular at $s_{+}$ to the SDC of $s_{+}$.} See Figs. \ref{fig.SDC+neg}, 
\ref{fig.SDC-neg} for more details. 
Therefore, in both cases, the asymptotic approximation of the integral is always based just on the contribution from the immediate neighbourhood of the saddle of interest.

The asymptotic approximation for the integral form of the solutions to the original real-variable problem, Eq.~(\ref{eq.inner}), then reads
\begin{multline}
 p_{\pm} =\Re\left[\int_C e^{\frac{\psi}{\varepsilon^*}} \right]\sim \Re\left[e^{\frac{\psi_{\pm}}{\varepsilon^*}}\int_{\mathrm{SDC_\pm}} e^{-\frac{1}{\varepsilon^*}\left(-\frac{\psi''_{\pm}}{2}\right)r^2}\dd r \right] \sim   \Re\left[e^{\frac{\psi_{\pm}}{\varepsilon^*}}\int_R e^{-\frac{1}{\varepsilon^*}\left(-\frac{\psi''_{\pm}}{2} e^{2 \ii \mu_{\pm}}\right)r^2}\dd r \right] \\= \sqrt{2 \pi \varepsilon^*} \left(-\psi''_{\pm}e^{2 \ii \mu_{\pm}}\right)^{-1/2} \exp\left(\frac{1}{\varepsilon^*} \Re \psi_{\pm}\right) \cos\left(\frac{1}{\varepsilon^*}\Im\psi_{\pm}\right).
\end{multline}
where the subscript corresponds to a solution passing through the $s_\pm$ saddle (two saddles) each generating two pairs of solutions (via its real and imaginary part as noted above; here it corresponds to replacing $\cos$ with $\sin$). 
%Similarly, taking the imaginary part would give the same expression, but replacing $\cos$ with $\sin$, and the second pair of solutions follows from the same asymptotics, but corresponding to the second saddle $s_{-}$, that is, evaluating the expression at $s_{-}$. 
In particular, when substituting the explicit expressions for the identified values of $\psi$ and its derivatives together with the tangent {\bl angle at the saddle,  $\mu_+= -0.0023\pi$ and $\mu_-= \pi/2+\mu_+ $, together  with $\kappa=\sqrt{-\beta\varepsilon^* z}/\alpha$ we have} 
\begin{align} \label{eq.inner.x=a.L<0.solu}
  p_{+}(x) &\sim \frac{\sqrt{\pi \varepsilon^*}}{2^{5/8}} \left(\frac{\alpha}{\beta}\right)^{1/4} (1+4 \kappa^2)^{1/8}\left[-\alpha e^{2 \ii\mu_{+}}\left(i\cos\varphi-\sin\varphi-\sqrt{1+4\kappa^2}[i\cos(3\varphi)-\sin(3\varphi)]\right)\right]^{-1/2}\nonumber\\ &\qquad\qquad\qquad\qquad\times\exp\left[\frac{(1+4\kappa^2)^{1/4}}{\sqrt{2}15}\frac{1}{\varepsilon^*}\frac{\alpha^{5/2}}{\beta^{3/2}}(-1+(1+4\kappa^2)^{1/2}-12\kappa^2)\sin\varphi\right]\nonumber\\ &\qquad\qquad\qquad\qquad\qquad\qquad\times\cos\left[\frac{(1+4\kappa^2)^{1/4}}{\sqrt{2}15}\frac{1}{\varepsilon^*}\frac{\alpha^{5/2}}{\beta^{3/2}}(-1+(1+4\kappa^2)^{1/2}-12\kappa^2)\sin\varphi\right]\\
  &\bl{\sim \frac{\sqrt{\pi \varepsilon^*}}{2^{5/8}} \left(\frac{\alpha}{\beta}\right)^{1/4}\left(1+\frac{\kappa^2}{4}\right)[-\alpha e^{2 \ii\mu_{+}}2\kappa(1+i \kappa)]^{-1/2}} \nonumber\\
  &\qquad\qquad\qquad\qquad\bl{\times\exp\left[-\frac{\alpha^{5/2} \sqrt{2}}{3 ~\varepsilon^* \beta^{3/2}} \kappa^2 \sin\varphi \right] \cos\left[\frac{\alpha^{5/2} \sqrt{2}}{3~\varepsilon^* \beta^{3/2}}(-10 \kappa^2) \sin\varphi \right]},\nonumber\\
  p_{-}(x) &\sim \frac{\sqrt{\pi \varepsilon^*}}{2^{5/8}} \left(\frac{\alpha}{\beta}\right)^{1/4} (1+4 \kappa^2)^{1/8}\left[-\alpha e^{2 \ii\mu_{-}}\left(i\cos\varphi+\sin\varphi-\sqrt{1+4\kappa^2}[i\cos(3\varphi)+\sin(3\varphi)]\right)\right]^{-1/2}\nonumber\\ &\qquad\qquad\qquad\qquad\times\exp\left[-\frac{(1+4\kappa^2)^{1/4}}{\sqrt{2}15}\frac{1}{\varepsilon^*}\frac{\alpha^{5/2}}{\beta^{3/2}}(-1+(1+4\kappa^2)^{1/2}-12\kappa^2)\sin\varphi\right]\nonumber\\
&\qquad\qquad\qquad\qquad\qquad\qquad\times\cos\left[\frac{(1+4\kappa^2)^{1/4}}{\sqrt{2}15}\frac{1}{\varepsilon^*}\frac{\alpha^{5/2}}{\beta^{3/2}}(-1+(1+4\kappa^2)^{1/2}-12\kappa^2)\sin\varphi\right]\\
&\bl{\sim \frac{\sqrt{\pi \varepsilon^*}}{2^{5/8}} \left(\frac{\alpha}{\beta}\right)^{1/4}\left(1+\frac{\kappa^2}{4}\right)[-\alpha e^{2 \ii\mu_{-}}2\kappa(1-i \kappa)]^{-1/2}} \nonumber\\
  &\qquad\qquad\qquad\qquad\bl{\times\exp\left[\frac{\alpha^{5/2} \sqrt{2}}{3 ~\varepsilon^* \beta^{3/2}} \kappa^2 \sin\varphi \right] \cos\left[\frac{\alpha^{5/2} \sqrt{2}}{3~\varepsilon^* \beta^{3/2}} \kappa^2 \sin\varphi \right]}.\nonumber
\end{align}

Note that all four solutions (the other pair is obtained by replacing $\cos$ with $\sin$ above) appear to be singular at the turning points. Specifically, the singularity here stems from the vanishing $\psi''$ at the saddles as $z\to 0^{-}$, while the remaining expressions are regular in this limit. We will therefore resolve the inner region further in \S \ref{sec_turning_pts}, noting that the inner solution identified here matches the WKBJ outer solution. Before this, however, we consider the additional turning point at $x = b$.

\subsubsection{Turning point $x=b$.} 
We now consider the second turning point at $x=b$. Using equivalent notation to above, with $\varepsilon z=x-b$, $\rho=r'(b)$, we obtain a similar expression for $p$, the solution to Eq.~\eqref{eq.innerp}, with a slightly modified argument of the exponential
\begin{equation*}
  \psi(s;z) = \ii\left[ -\varepsilon^* z s - \frac{\alpha}{3} s^3+\frac{\beta}{5}s^5\right].
\end{equation*}
Hence, the corresponding saddles read:
\begin{align}\label{eq.saddles_x=b}
s_{+}&=\left(\frac{\alpha}{2\beta}\right)^{1/2}\left[1+\left(1+4\frac{\beta}{\alpha^2}\varepsilon^* z\right)^{1/2}\right]^{1/2},%\quad s_{-+}=-s_{+},\\
 & s_{-}=\left(\frac{\alpha}{2\beta}\right)^{1/2}\left[1-\left(1+4\frac{\beta}{\alpha^2}\varepsilon^* z\right)^{1/2}\right]^{1/2},%\quad s_{--}=-s_{-}.
\end{align}
and there is another pair with the opposite signs. 
Hence, $s_{+}$ is always real for small $\varepsilon^* z$, while $s_{-}$ is real for $z<0$ and purely imaginary for $z>0$.

Again, the tangent of the SDC at the saddle is equal to the value of $\mathrm{arg}\left[(-\psi'')^{-1/2}\right]$ evaluated there. Near $x=b$, this gives for $z<0$
\begin{equation*}
  (\psi''(s_{+}))^{-1/2}\propto (-\ii)^{-1/2},
\end{equation*}
where the constant of proportionality is real, so the angle of the SDC at $s_{+}$ is $\pi/4$. The same procedure yields the tangent of the SDC at $s_{-}$ to be $3\pi/4$. For $z>0$ (with one real saddle and the other purely imaginary) the angle is constant and equal to $0$ for both saddles. We note that the asymptotes of the SDCs are the same for $x=b$ as for $x=a$ since both have $\psi(s;z)\sim \ii s^{5} / 5$ for large $|s|$. Hence, the asymptotes are $\mathrm{arg}(s)= \pi/10 + 2k \pi/5$ for $k\in\{0,1,\ldots,4\}$.

\paragraph{Real case $z<0$.} We again start with the simpler real case where $z<0$. When $0<-\varepsilon^* z=-\rho(x-b)\ll 1$, the results above give the following asymptotic approximations of the solutions to the original real problem, Eq.~%\todo{ckontroluj znamenka, vyraz.. rozvoj nesedi p+ na to nize.. ani pro sin..}
(\ref{eq.inner})
\begin{subequations}\label{eq.inner.x=b.L<0.solu}
\begin{align} 
  p_{+}(x) &\sim \sqrt{\pi \varepsilon^*} \frac{(2\beta)^{1/4}}{\alpha^{3/4}} \frac{1}{(1-4\kappa^2)^{1/4}}\frac{1}{(1+\sqrt{1-4 \kappa^2})^{1/4}}\nonumber\\ &\qquad\qquad\qquad\qquad\qquad\qquad\times\cos\left[\frac{1}{2}\frac{\sqrt{2}}{15}\frac{\alpha^{5/2}}{\varepsilon^*\beta^{3/2}}(1+\sqrt{1-4 \kappa^2}-12\kappa^2)(1+\sqrt{1-4\kappa^2})^{1/2}\right],\\
  &\bl{\sim \sqrt{\pi \varepsilon^*} \frac{\beta^{1/4}}{\alpha^{3/4}} \left(1+\frac{5 \kappa^2}{4}\right)\cos\left[\frac{1}{15}\frac{\alpha^{5/2}}{\varepsilon^*\beta^{3/2}}(2-15 \kappa^2)\right]},\nonumber\\
    p_{-}(x) &\sim \sqrt{\pi \varepsilon^*} \frac{(2\beta)^{1/4}}{\alpha^{3/4}} \frac{1}{(1-4\kappa^2)^{1/4}}\frac{1}{(1-\sqrt{1-4 \kappa^2})^{1/4}} \nonumber\\ &\qquad\qquad\qquad\qquad\qquad\qquad\times\cos\left[\frac{1}{2}\frac{\sqrt{2}}{15}\frac{\alpha^{5/2}}{\varepsilon^*\beta^{3/2}}(1-\sqrt{1-4 \kappa^2}-12\kappa^2)(1-\sqrt{1-4\kappa^2})^{1/2}\right],\\
     &\bl{\sim \sqrt{\pi \varepsilon^*} \frac{\beta^{1/4}}{\alpha^{3/4}} \kappa^{-1/2}\cos\left[-\frac{2}{3}\frac{\alpha^{5/2}}{\varepsilon^*\beta^{3/2}}\kappa^3\right]},\nonumber
\end{align}
\end{subequations}
where we use $\kappa=\sqrt{\beta\rho(b-x)}/\alpha>0$ and where  the subscript corresponds to the chosen saddle approximation. There is an additional pair of solutions when $\cos$ is replaced with $\sin$. Note that these solutions are purely oscillatory due to the fact that $\Re(\psi)(s_\pm)=0$.

We note that the inner solutions around the turning point at $x=b$ behave differently to the inner solutions around the turning point at $x=a$. In particular, the solutions corresponding to real and imaginary parts of $p_{+}$ are regular as $\kappa\to 0^{-}$
%\begin{equation*}
%  p_{+}(x)\sim \frac{2}{3}\sqrt{\frac{\pi}{\varepsilon^*}} \frac{\alpha^{7/4}}{\beta^{5/4}} \kappa^{5/2}= \frac{2}{3}\sqrt{\frac{\pi}{\varepsilon^*}} \frac{1}{\alpha^{3/4}} [\rho(-\Lambda)]^{5/2},
%\end{equation*}
and hence this asymptotic solution gives us an approximation valid across the turning point $x=b$. The second pair of solutions, corresponding to $p_{-}$, does have a singular amplitude as we approach the turning point, $\kappa \to 0^{-}$. However, since the frequency of the oscillations also vanishes in this limit, the $\sin$ solution is regular as we approach the turning point. That is, for small $\kappa>0$ we have
\begin{equation*}
p_{-}(x) \sim -\sqrt{\pi \varepsilon^*} \frac{\beta^{1/4}}{\alpha^{3/4}} \kappa^{-1/2} \sin\left[\frac{10}{15}\frac{\alpha^{5/2}}{\varepsilon^*\beta^{3/2}}\kappa^3\right]\sim -\frac{10}{15}\sqrt{\frac{\pi}{\varepsilon^*}}\frac{\alpha^{7/4}}{\beta^{5/4}} \kappa^{5/2}
\end{equation*}
as $\kappa=\sqrt{\beta\rho(b-x)}/\alpha\to 0^{+}$. The $\cos$ solution does not have this property however, and is singular at the turning point. Therefore, we will resolve this $\cos$ solution in the imaginary case $z>0$ below.

%Imag case, str 14
\paragraph{Imaginary case $z>0$.} 
Finally, we consider the case $z>0$ which corresponds to the partially complex case. First, we rewrite the saddles, Eq.~\eqref{eq.saddles_x=b}, so that their real and imaginary part are explicit:
\begin{align}\label{eq.saddles_x=b_imag}
s_{+}&=\left(\frac{\alpha}{2\beta}\right)^{1/2}\left[1+\left(1+4\frac{4\beta}{\alpha^2}\varepsilon^* z\right)^{1/2}\right]^{1/2}=\left(\frac{\alpha}{2\beta}\right)^{1/2}\left[1+\sqrt{1+4\kappa^2}\right]^{1/2}\in\mathbb{R},\\
s_{-}&= \ii\left(\frac{\alpha}{2\beta}\right)^{1/2}\left[-1+\left(1+4\frac{4\beta}{\alpha^2}\varepsilon^* z\right)^{1/2}\right]^{1/2}=i\left(\frac{\alpha}{2\beta}\right)^{1/2}\left[-1+\sqrt{1+4\kappa^2}\right]^{1/2}\in i\mathbb{R},
\end{align}
where $\kappa=(\beta\varepsilon^* z)^{1/2}/\alpha$. We see that, at leading order, $s_{+}$ generates an oscillatory solution, while $s_{-}$ generates a purely exponential solution.%\todo{there are several kappas throughout.. should we distinguish them? I opted for stating explicitly what they are whenever there is a change but...}

Recalling that we only require consideration of the $\cos$ case here ($p_{-}$ with a purely imaginary saddle $s_{-}\in i\mathbb{R}$), the same procedure as above for $x=a$ yields the following asymptotic result:
\bl{
\begin{multline} \label{eq.inner.x=b.L>0.solu}
  p_{-}(x) \sim - \sqrt{\pi \varepsilon^*} \frac{(2 \beta)^{1/4}}{\alpha^{3/4}}\left[(1+4\kappa^2)\left(1-\sqrt{1+4\kappa^2}\right)^3\right]^{-1/4}\\
\times\exp\left[-\frac{1}{15\sqrt{2}}\frac{1}{\varepsilon^*}\frac{\alpha^{5/2}}{\beta^{3/2}}\left(-1+\sqrt{1+4\kappa^2}\right)^{1/2}(1-\sqrt{1+4\kappa^2}+12\kappa^2)\right],
\end{multline}
where we note that this is a purely exponential solution due to the fact that $\Im(\psi)(s_-)=0$ and the tangent of the SDC at the saddle $s_{-}$ is zero.  This solution allows the following approximation
\begin{equation*}
  p_{-}(x) \sim - \frac{\sqrt{\pi \varepsilon^*}}{(\alpha \varepsilon^* z)^{1/4}} \exp\left[-\frac{2 (\varepsilon^* z)^{3/2}}{3\sqrt{\alpha} \varepsilon^*}\right].
\end{equation*}
Finally, we note that the solution obtained for $z>0$ in Eq.~\eqref{eq.inner.x=b.L>0.solu} is in line with what we have above for the negative case $z<0$. There, we see that the solution corresponding to $s_-$ has vanishing frequency as we approach the turning point $x=b$ and this property remains true from either side of the turning point. 
}

%%%%%%%%%%%%%%%%%%%%%%%%%%%%%%%%%%% coalescing saddles %%%%%%%%%%%%%%%%%
\section{Behaviour near turning points due to coalescing saddles}\label{sec_turning_pts}
In the preceding section, the asymptotic approximations of the integral form of the solution generated apparent singularities near turning points. This singular behaviour is not the correct behaviour of the actual system, and therefore we have not yet provided a complete description of the solution behaviour across the turning points. The reason for the apparent singularities in the integral form lies in the coalescing of the two saddles as the turning points are approached. As a result, the quadratic approximation of the argument of the exponential is not sufficient, since this flattens out once the two saddles coalesce and simply retaining higher-order terms is not sufficient. A more subtle approach is required, which is due to Chester, Friedman, and Ursell \cite{chester1957extension} when there is a pair of saddles close to coalescence. See also \cite{temme2014asymptotic, olver1997asymptotics, wong2001asymptotic} for comprehensive discussions on the need for this approach.% Other nice literature on this topics are \cite but note the different sign notation in the definition of the Airy functions which in turn changes the corresponding choice of SDC.

In \cite{chester1957extension}, Chester et al.~devised a method for improving the asymptotic accuracy of a third-order {\bl polynomial -- the lowest order with the property of two coalescing saddles -- in the exponent of an integrand subjected to integral asymptotics.}  Their method requires finding a suitable (local) transformation in the complex plane so that the two coalescing saddles are mapped onto the two coalescing saddles of the cubic. In turn, the integral involving a cubic polynomial can be asymptotically approximated using the integral representation of Airy functions (by a recurrence sequence involving Airy functions and their derivatives). This last step also occurs via a transformation in the complex plane, and significant care is required to choose the appropriate branches of roots in the complex plane where the Airy curve is selected.

\begin{figure}
\centering
   \includegraphics[width=.3\linewidth]{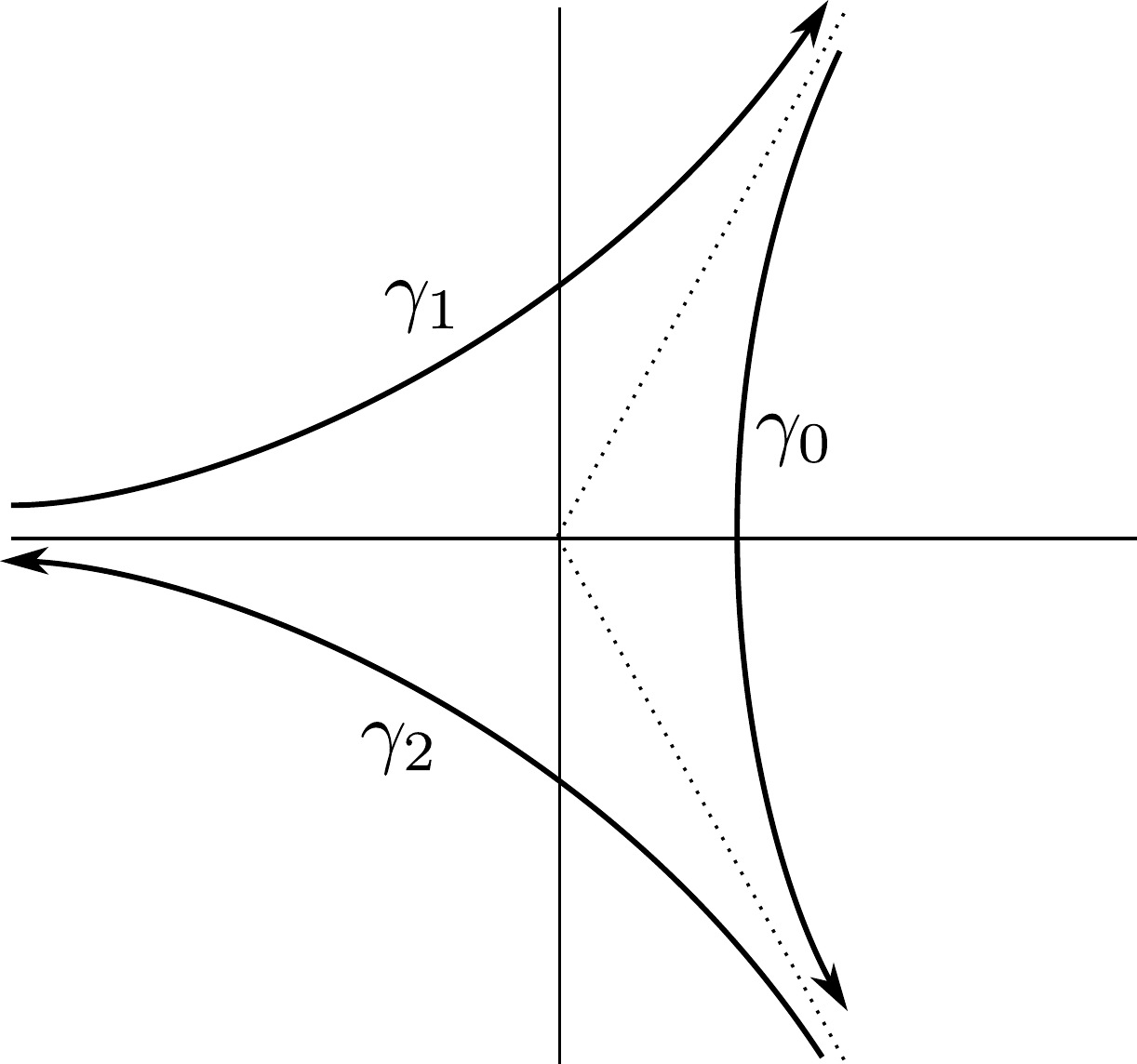}
\caption{\label{fig.CubicCurvesAi} Contours generating Airy functions. Contour $\gamma_{0}$ corresponds to $\mathrm{Ai}$, while the remaining two generate a linear combination of $\mathrm{Ai}$, $\mathrm{Bi}$.
}
\end{figure}

Let us denote the integral involving the cubic polynomial with coalescing saddles as
\begin{equation}
  \label{eq:Int-CubicCoalesc}
  I(\varepsilon^{*};\eta) = \int_{C} f(t) \exp\left(\frac{\Xi(t)}{\varepsilon^{*}}\right)\dd t,\quad \Xi(t) = \frac{1}{3}t^{3}-\eta t+A,
\end{equation}
with two coalescing saddles $t_{\pm}=\pm \eta^{1/2}$ as $\eta\to0^{+}$, $0 < \varepsilon^{*} \ll 1$, and $f$ analytic. To have a convergent integral, the asymptotes of the integration curve $C$ need to be the cubic roots of minus unity, as shown in Figure \ref{fig.CubicCurvesAi}.

Next, we recall the integral representation of Airy functions,
\begin{equation}
  \label{eq.Airy}
  \mathrm{Ai}(\eta)=\frac{1}{2\pi i} \int_{\gamma_{0}} \exp\left(\frac{1}{3} t^{3}-\eta t\right)\dd t, \quad \eta\in\mathbb{C}.
\end{equation}
The dominant term is $t^{3}/3$ for $t\to\infty$ and hence the asymptotes of the integration curve are again the cubic roots of minus unity. The saddles occur at $t = \pm \eta^{1/2}$ and the Airy function $\mathrm{Ai}$ corresponds to the curve $\gamma_{0}$ with the orientation indicated in Figure \ref{fig.CubicCurvesAi}. Other choices of integration curves give some linear combination of $\mathrm{Ai}$ and $\mathrm{Bi}$. The asymptotics of Airy functions for large arguments can be obtained via a deformation of the integration curve to the SDC. One may rescale the integrand so that the saddles are normalised and fixed by a transformation $t=\eta^{1/2}z$. If $\Im (\eta^{1/2}) \neq 0$ (the case in which we are interested), this transformation also rotates the integration curve.

In \cite{chester1957extension}, the authors considered exactly this rescaled form of the Airy function:
\begin{equation}
  \label{eq:AiryRescaled2pm1}
  \mathrm{Ai}(\eta) = \frac{\eta^{1/2}}{2 \pi i} \int_{\tilde{\gamma}_{0}} \exp\left(\eta^{3/2}\left(\frac{1}{3}z^{3}-z\right)\right)\dd z,
\end{equation}
which keeps the saddles at a fixed distance while the asymptotic parameter $\eta$ varies. Here, $\tilde{\gamma}_{0}$ is the Airy curve $\gamma_{0}$ rotated (and rescaled) by $\mathrm{arg}(\eta^{{1/2}})$ %\textcolor{blue}{\sout{counter}}
clockwise. \bl{Note that the curve $\gamma_0$ is the SDC only for $|\arg(\eta^{1/2})|<\pi/3$ and that the $z = 1$ saddle governs the asymptotic behaviour of the Airy function $\mathrm{Ai}(\eta)$ in the Airy integral in the rescaled form, Eq.~\eqref{eq:AiryRescaled2pm1}.}

Once we have specified the integration curve, we can approximate the integral $I(\varepsilon^{*};\eta)$ with the coalescing saddles from Eq.~\eqref{eq:Int-CubicCoalesc}, noting that \cite[Chap 23]{temme2014asymptotic}, \cite[Chap 9]{olver1997asymptotics}, \cite{chester1957extension,wong2001asymptotic} we have
%\begin{multline} \label{eq:Int-CubicCoalescAsympt}
  %\frac{1}{2\pi i} \int_{\gamma_0}f(t) \exp\left[\frac{1}{\epsilon}\left(\frac{1}{3} t^3 - \eta t\right)\right] \dd t \sim \epsilon^{1/3}\left[\frac{1}{2}\left(f(\eta^{1/2})+f(-\eta^{1/2})\right)+\Ord{\epsilon}\right] \mathrm{Ai}\left(\eta \epsilon^{-2/3}\right)\\-\epsilon^{2/3}\left[\frac{1}{2}\frac{1}{\eta^{1/2}}\left(f(\eta^{1/2})-f(-\eta^{1/2})\right)+\Ord{\epsilon}\right] \mathrm{Ai}'\left(\eta \epsilon^{-2/3}\right),
%\end{multline}
\bl{
\begin{equation} \label{eq:Int-CubicCoalescAsympt}
  \frac{1}{2\pi i} \int_{\gamma_0}f(t) \exp\left[\frac{1}{\epsilon}\left(\frac{1}{3} t^3 - \eta t\right)\right] \dd t \sim \epsilon^{1/3}\left[\frac{1}{2}\left(f(\eta^{1/2})+f(-\eta^{1/2})\right)+\Ord{\epsilon}\right] \mathrm{Ai}\left(\eta \epsilon^{-2/3}\right)+\O(\epsilon^{2/3}),
\end{equation}
}
as $\epsilon\to 0^+$. We label the coefficient from the above expression as
\begin{equation}
    \label{eq:A0B0}
    A_{0}=\frac{1}{2}\left(f(\eta^{1/2})+f(-\eta^{1/2})\right).%,\quad %B_{0}=\frac{1}{2}\frac{1}{\eta^{1/2}}\left(f(\eta^{1/2})-f(-\eta^{1/2})\right).
\end{equation}

In order to use the asymptotic approximations of $\mathrm{Ai}$ correctly, we need to understand how the scaling $\eta^{1/2}$ affects the choice of signs and branches of roots in the complex plane. Specifically, it transforms the integration curve $\gamma_{j}$ to $\tilde{\gamma}_{j}$ via a scaling and a rotation of angle $\mathrm{arg}(\eta^{{1/2}})$, where the two branches of the square root $\eta^{1/2}$ correspond to a sign change of this rotation. Hence, this rescaling and rotation may result in the inability to deform the $\mathrm{Ai}$ integration curve $\gamma_{0}$ into the SDC for large $\eta\in\mathbb{C}$ in magnitude, where instead $\gamma_{1}$ or $\gamma_{2}$ would have to be chosen. Note that choosing the integration curve $\gamma_{j}$ in the definition of the Airy integral as depicted in Fig.~\ref{fig.CubicCurvesAi} results in the expression
\begin{equation*}
  \frac{1}{2\pi i} \int_{\gamma_{j}}e^{\frac{1}{3} t^{3}-\eta t}\dd t=\omega^j\mathrm{Ai}(\omega^j\eta),\quad \omega=e^{2\pi i/3},
\end{equation*}
as a suitable transformation reveals.

%$\pi/2$ counter clockwise. %-pi/2+pi..arg of complex numbers is the angle taken counter-clockwise from real axis % ... fig nahore na str 51
%In addition, we need to determine which of the $\pm \eta^{1/2}$ saddles is mapped onto which of the rescaled $\pm 1$ saddles. For this we need to identify the suitable value of $\eta$ first but also the SDC through the two saddles $s_{++},~s_{+-}$. We start with the SDC

Therefore, to employ the method of Chester et al.~we need to: (i) find a suitable (local nonlinear) transform that casts the integral form of the solution (\ref{eq.innerp}) into the form of the cubic polynomial exponent (\ref{eq:Int-CubicCoalesc}). That is, we must identify the appropriate values of $A,~\eta$ in Eq.~\eqref{eq:Int-CubicCoalesc} and the integration curve $C$, as well as which saddle in the coalescing pair maps to which of the $\pm 1$ saddles of the rescaled cubic; and (ii) identify the asymptotics of the sought solution via \eqref{eq:Int-CubicCoalescAsympt} by identifying the function $f$ from the (local) nonlinear transform.

To this end, a cubic transformation is used in order that the two coalescing saddles are mapped onto the two saddles of the cubic. In particular, we consider
\begin{equation*}
  \Xi(t) = \frac{1}{3} t^3-\eta t+A,
\end{equation*}
and we look for the values of parameters $A,~\eta$ such that the two saddles of the cubic, that is $t_\pm=\pm\eta^{{1/2}}$, match the two coalescing saddles $s_{+},~s_{-}$. Hence, we look for a transformation $s\to t$ such that
\begin{equation*}
  \psi(s)=\Xi(t) \mbox{ at the two matching saddles.}
\end{equation*}

To choose the correct branches, we define for uniqueness the two saddles $t_{\pm}$ as
  \begin{equation}\label{eq:tpm_saddles}
t_{\pm} = \pm e^{i\tau}|\eta|^{1/2},\quad \tau\in[0,\pi),
\end{equation}
and hence $t_{+}$ is chosen for all $\eta$ in such a way that its imaginary part is positive. Thus we have $\eta=e^{2 \ii\tau}|\eta|$ and we may write
\begin{equation}
    \label{eq:A0B0corr}
    A_{0}=\frac{1}{2}\left(f(t_{+})+f(t_{-})\right).%,\quad B_{0}=\frac{1}{2}\frac{1}{t_{+}}\left(f(t_{+})-f(t_{-})\right).
\end{equation}
Now, depending on which saddle is mapped to which, we have either $\psi(s_{\pm})=\Xi(t_{\pm})$ or $\psi(s_{\pm})=\Xi(t_{\mp})$. Since we have
\begin{equation*}
\Xi(t_\pm) = \mp \frac{2}{3} e^{3 \ii\tau}|\eta|^{3/2}+A,
\end{equation*}
in either case, we note that in the case $s_{+} \mapsto t_{+}$, we require that
\begin{align*}
  A&=\frac{1}{2}\left(\psi(s_{+})+\psi(s_{-})\right),\\
  \frac{4}{3}e^{3 \ii\tau}|\eta|^{3/2}&=\psi(s_{-})-\psi(s_{+})\equiv e^{i\sigma}|\Sigma|,
\end{align*}
where we have introduced the argument ($\sigma$) and absolute value ($|\Sigma|$) of the difference $\psi(s_{-})-\psi(s_{+})$.

In the case $s_{+} \mapsto t_{-}$, the expression for $A$ is as above (due to its symmetry), while the second expression changes sign. Specifically, for the case $s_{+} \mapsto t_{-}$, we have
\begin{equation*}
    \frac{4}{3}e^{i3\tau}|\eta|^{3/2}=- e^{i\sigma}|\Sigma|.
\end{equation*}

Hence, we always have
  \begin{equation}\label{eq:etaabs}
    |\eta|=\left(\frac{3}{4}|\Sigma|\right)^{2/3},
  \end{equation}
but the reconstruction of the angle $\tau$ depends on the mapping of saddles. For the case $s_{+}\to t_{+}$ we have
\begin{equation}\label{eq:tau_sigma:s++Tot+}
    \tau =(\sigma+2 n \pi)/3, \quad n\in\{0,1,2\},
\end{equation}
while for the case $s_{+}\to t_{-}$ we have
\begin{equation}\label{eq:tau_sigma:s++Tot-}
    \tau =(\sigma+(2 n +1)\pi)/3, \quad n\in\{0,1,2\}.
\end{equation}
\bl{Since $\tau \in [0,\pi)$, when $s_{+}\to t_{+}$ we have 
\begin{align*}
    &\sigma\in[0,\pi) \implies \tau=\sigma/3 \mbox{ or } \tau=(\sigma+2\pi)/3,\\
    &\sigma\in[\pi,2 \pi) \implies \tau=\sigma/3.
\end{align*}
Similarly, for $s_{+}\to t_{-}$ we have 
\begin{align*}
    &\sigma\in[0,\pi) \implies \tau=(\sigma+\pi)/3,\\
    &\sigma\in[\pi,2 \pi) \implies \tau=(\sigma+\pi)/3 \mbox{ or } \tau=(\sigma-\pi)/3.
\end{align*}
In short, it is crucial to assess which saddle $s_\pm$ maps onto which $t_\pm$. This can be done by inspecting the landscapes of $\psi(s)$ and $\Xi(t)$, and from the local transformation $s(t)$. 
}

%Next, one can show from the asymptotics of the Airy function for a large argument that the $\tilde{\gamma}_{0}$ contour can be deformed to the SDC passing through the normalised saddle $+1$ if $\mathrm{arg}(t_{+})=\tau\in[0,\pi/3)$ where the upper bound $\pi/3$ is known as the Stokes line (both saddles are then connected by a single SDC), see \cite[Chapter 4]{temme2014asymptotic} for recalling the asymptotics of $\mathrm{Ai}$. Conversely, $\tilde{\gamma}_{0}$ contour can be deformed to the SDC passing through the normalised saddle \textcolor{blue}{$+1$} if $\mathrm{arg}(t_{-})\in(-\pi/3,0]$. \textcolor{blue}{The former is via the mapping $t_+\to+1$ while the latter via $t_-\to +1$.} Thus, we observe from relations \eqref{eq:tau_sigma:s++Tot+}, \eqref{eq:tau_sigma:s++Tot-} that the sought transformation takes the route $s_{+}\to t_{+}$ if $\sigma\in(0,\pi)$ and $s_{+}\to t_{-}$ if $\sigma\in(-\pi,0)$.

This understanding leads us to the following workflow: the value of $\varepsilon^* z$ uniquely determines both $\sigma$ and $\Sigma$, which in turn specifies $|\eta|$, though the argument $\tau$ still has three distinct possibilities depending on the saddle pairing. %We have
%\begin{equation} \label{eq.tauCond}
  %\tau = (\sigma+2 n \pi)/3\in(0,\pi/3) \mbox{ for $s_+\to t_+$} \quad \mbox{  or  } \quad  \tau= (\sigma+2 n \pi)/3\in(2\pi/3,\pi) \mbox{ for $s_+\to t_-$},\quad n\in\{0,1,2\}.
%\end{equation}
%Note that there is a unique natural $n\in\{0,1,2\}$ which satisfies this last expression for any given $\sigma$. Therefore, both the value of the argument $\tau$ and the saddle mapping (whether $s_{+}$ is mapped onto $t_{+}$ or $t_{-}$) is uniquely determined. 
In addition, this pairing then affects the expression for the $A_0$ %, ~B_0$ 
 coefficients via the transformation function $f(t)$. 

Hence, the last step is to identify the suitable function $f(t)$ corresponding to the transformation. Recalling that we use a local transformation to rewrite the integral form of solution in the cubic form
\begin{equation*}
 \int_{C} \exp\left(\frac{\psi(s)}{\varepsilon^{*}}\right)\dd s \sim \int_{\gamma_0} f(t) \exp\left(\frac{\Xi(t)}{\varepsilon^{*}}\right)\dd t=\exp\left(\frac{A}{\varepsilon^{*}}\right)\int_{\gamma_0} f(t) \exp\left(\frac{t^{3}/3-\eta t}{\varepsilon^{*}}
 \right) \dd t,
\end{equation*}
we can determine $f(t)=\dd s/\dd t$ from the implicit relation $s=s(t)$, which follows from the expansions of both $\psi$ and $\Xi$ about the corresponding saddles. For example, in the case $s_{+}\to t_{+}$ we require $\Xi(t) \sim \psi(s)$ where
\begin{align*}
\psi(s)&=\psi(s_{+})+ \frac{\psi''(s_{+})}{2}(s-s_{+})^2+\frac{\psi'''(s_{+})}{6} (s-s_{+})^3 + \Ord{s-s_{+}}^4\\
    \Xi(t) &=\Xi(t_+) + \frac{\Xi''(t_+)}{2} (t-t_+)^2+\frac{\Xi'''(t_+)}{6} (t-t_+)^3 + \Ord{t-t_+}^4,
\end{align*}
% \begin{align*}
%   \Xi(t)&=\psi(s)\\
%         &=\psi(s_{+})+\psi''(s_{+}) \frac{1}{2}(s-s_{+})^2+\psi'''(s_{+})\frac{1}{6} (s-s_{+})^3 + \Ord{s-s_{+}}^4\\
%   &=\Xi(t_+) + \Xi'(t_+) \frac{1}{2} (t-t_+)^2+\Xi'''(t_+)\frac{1}{6} (t-t_+)^3 + \Ord{t-t_+}^4,
% \end{align*}
and hence we identify $s(t)$ near $t=t_+$ as
\begin{equation*}
  s(t) = s(t_+) \pm \left(\frac{\Xi''}{\psi''}\right)^{1/2}(t-t_+)\pm\frac{\Xi'''\mp\psi'''\left(\frac{\Xi''}{\psi''}\right)^{3/2}}{3 (\Xi'' \psi'')^{1/2}} (t-t_+)^2+\Ord{t-t_+}^3,
\end{equation*}
where $\Xi$ and $\psi$ (and their derivatives) are evaluated at $t_+$ and $s_{+}=s(t_{+})$ respectively. An analogous expression is valid for the case $s_{+}\to t_{-}$.

For the leading-order asymptotic approximation of the integral it is enough to assess $A_{0}$ %$,~B_{0}$
from \eqref{eq:A0B0corr} and hence we only need to evaluate $f(t_{\pm})$, that is
\begin{equation*}
  f(t_{\pm})=\pm \left(\frac{\Xi''(t_{\pm})}{\psi''(s(t_{\pm}))}\right)^{1/2},
\end{equation*}
where we need to assess the sign to complete the transformation. %, we need to assess the sign in front of, that is the branch of, the square root in the linear term.
%Recalling the relation for $B_{0}$ in Eq.~(\ref{eq:A0B0corr}), we infer that for $B_{0}$ not to blowup, we require that
%$$\textcolor{blue}{\mathrm{arg}(f(t_{-}))=-\mathrm{arg}(f(t_{+}))}$$
%noting that already Chester et al \cite{chester1957extension} showed that there is a local one-to-one analytic transformation to the Airy type integral. 
Therefore, it remains to determine the branch of $f(t_{+})$ which follows from the rotation of the asymptotes (Stokes lines) from the $\Xi(t)$ polynomial to the $\psi(s)$ polynomial so that the direction of integration along the curve does not change. \bl{The final information is the mapping between saddles. To this end, we recall that the saddle governing the Airy function asymptotics occurs at $+1$  in Eqn.\eqref{eq:AiryRescaled2pm1}, with the {\it lower} value of the real part of the exponential  on comparing the two saddles, see \cite[Chapter 4]{temme2014asymptotic}. Hence we choose the mapping  of the saddles such that  $s_+$ maps to the saddle in $t-$space that possesses the {\it lower} real part on examining the exponent $\Xi(t)$  at the saddle points $t_+$ and $t_-$.}

\paragraph{Turning point $x=a$.} 
\bl{
First, we consider the turning point $x=a$, where we note that the saddles $s_\pm$ are real for $z>0$ and purely imaginary for $z<0$. Next, we see from the real part of $\psi$, Eq.~\eqref{eq.ReImPsi-re}, that $\Re\psi(s_\pm)=0$ for $z>0$ and hence we cannot discriminate the saddle mapping based on this information. However, for the imaginary case, $z<0$, we have that $0<-\Re(\psi_+)=\Re(\psi_-)$ as shown in Eqn. \eqref{eq.Rex=aImagCase}. Therefore, when applying the method of coalescing saddles we select the imaginary case $z<0$ where $s_+\mapsto t_+$. 
}

\bl{
Next, for the approximation near $x=a$ we know that $\Im\psi(s_+)=\Im\psi(s_-)$, see Eq.~(\ref{eq.Imx=aImagCase}), and hence 
\begin{equation*}
  \mathrm{arg}(\psi(s_{-})-\psi(s_{+})) = \arg(2\Re\psi_-) = 0 = \sigma,
\end{equation*}
while $|\Sigma|=2\Re\psi_-$. 
Thence either $\tau=0$ or $\tau=2\pi/3$. However, the corresponding SDC cannot be deformed to the curve $\gamma_0$ when $\tau=2\pi/3$, as discussed above. Therefore, we must have $\tau = 0$, and hence we find that the saddle $s_{{+}} \mapsto t_{+}$ with
\begin{align*}
  \tau&=0,\\ \eta&=|\eta|=\left(\frac{3}{2} \Re\psi_-\right)^{2/3}\sim -(2 \alpha)^{-1/3}\varepsilon^* z\left(1+\frac{\beta}{5 \alpha^2} \varepsilon^* z+\O((\varepsilon^* z)^2)\right),\\
  A&=i \Im\psi_+ \sim \ii\left(\frac{2\alpha}{ \beta}\right)^{1/2}\left(\frac{1}{15}\frac{\alpha^2}{\beta}-\frac{\varepsilon^* z}{2}+\O((\varepsilon^* z)^2)\right).
\end{align*}
}

\bl{
Next, we determine $f(t)$ from the transformation $s(t)$. To this end, we evaluate the absolute value and argument of $\psi''(s_+)$. We have
\begin{equation*}
 \psi''(s_\pm) \sim \mp (2 \alpha)^{1/2} (-\varepsilon^* z)^{1/2}\left(2+\frac{\beta}{\alpha^2}(-\varepsilon^* z)\right) - \ii\left(2 \frac{\beta}{\alpha}\right)^{1/2} (-\varepsilon^* z)\left(2-\frac{\beta}{\alpha^2}(-\varepsilon^* z)\right)
\end{equation*}
and hence
\begin{equation*}
    |\psi''(s_\pm)|\sim 2 \sqrt{2}\alpha^{3/2} \beta^{-1/2} \kappa = 2 \sqrt{2 \alpha} (-\varepsilon^* z)^{1/2},
\end{equation*}
together with
\begin{equation*}
\arg(\psi''(s_\pm))\sim\pm \frac{\sqrt{\beta(-\varepsilon^* z)}}{\alpha}\left(1-\frac{4}{3}\frac{\beta}{\alpha^2}(-\varepsilon^* z)\right),  
\end{equation*}
as $\varepsilon^* z\to0^-$.
}

\bl{
It follows that
\begin{equation*}
  |f(t_\pm)| =\frac{|\eta|^{1/4}}{|\psi''(s_\pm)|^{1/2}}\sim \frac{1}{(2\alpha)^{1/3}}\left(1+\frac{11}{20}\frac{\beta}{\alpha^2}\varepsilon^* z+\O((\varepsilon^* z)^2)\right),
\end{equation*}
while for the argument of $f(t_{\pm})$ we have 
\begin{equation*}
  \arg(f(t_{\pm})) = \arg\left(\frac{\pm e^{i\tau}}{\psi''(s_\pm)}\right)^{1/2}=\mp\frac{1}{2}\arg\left(\psi''(s_\pm)\right)+n \pi, \quad n\in\{0,1\}
\end{equation*}
where $n$ chooses the branch of the square root. Finally, the sought coefficient of the Airy function reads
\begin{equation}
     \Re(A_0) = \frac{1}{2}\Re(f(t_+)+f(t_-)) = \frac{1}{2}|f(t_+)|\cos\left(\arg(f(t_{+}))\right) \\ \sim (2\alpha)^{-1/3} \left(1+\frac{27}{40}\frac{\beta}{\alpha^2}\varepsilon^* z+\O((\varepsilon^* z)^2)\right),
\end{equation}
irrespective of the chosen branch in $\arg(f(t_\pm))$. That is, for any $n\in\{0,1\}$, we have $\Im(A_0)=0$ and hence $A_0=\Re(A_0)$. Finally, we note that the same asymptotic approximations are obtained in the real case, $z>0$, where now the saddle mapping is $s_+\mapsto t_-$ which in turn selects $\tau=\pi/2$. For explicit expressions, we refer the reader to \cite{krause2024pattern}.
}

Therefore, the coalescing method approximates the behaviour of the (integral) solution across the turning point to be the real or imaginary part of the following expression
\begin{equation} \label{eq.TurningPointSolu_x=a}
 p=\int_{C} \exp\left(\frac{\psi(s)}{\varepsilon^{*}}\right)\dd s = \exp \left(\frac{A}{\varepsilon^{*}}\right)\int_{\gamma_0} f(t)
 \exp \left(\frac{t^{3}/3-\eta t}{\varepsilon^{*}} \right)\dd t \sim 2\pi \ii~e^{A/\varepsilon^*}%\left[
 (\varepsilon^*)^{1/3} \Re(A_0) \mathrm{Ai}\left(\eta (\varepsilon^*)^{-2/3}\right),%-(\varepsilon^*)^{2/3} B_0 \mathrm{Ai}'\left(\eta (\varepsilon^*)^{-2/3}\right)\right],
\end{equation} 
where all the coefficients are now uniquely identified. \bl{In particular, the imaginary part of \eqref{eq.TurningPointSolu_x=a} gives the cosine  oscillatory solution
\begin{equation}\label{eq.TurningPointSolu_x=a_cos}
    \Im(p(y))\sim 2 \pi \left(\frac{\varepsilon^*}{2\alpha}\right)^{1/3} \cos\left(\left(\frac{\alpha}{2 \beta}\right)^{1/2} \frac{x-a}{\varepsilon}+\frac{1}{\varepsilon^*}\left(\frac{\alpha}{2 \beta}\right)^{1/2}\frac{2}{15}\frac{\alpha^2}{\beta}\right) \mathrm{Ai}\left(-\left(\frac{\rho}{2\alpha}\right)^{1/3} \frac{x-a}{\varepsilon^{2/3}}\right),
\end{equation}
where we used the fact that $\eta\in \mathbb{R}$ and hence the Airy function in  Eq.~\eqref{eq:Int-CubicCoalescAsympt} takes real values.}

\paragraph{Turning point $x=b$.}

The final step is to return to the turning point $x=b$. Here, it is only the $p_{-}$ case which requires the finer approach of coalescing saddles to resolve the transition across the turning point. We first note that a different pair of saddles coalesce here, specifically $s_{{+-}}$ and $s_{--}=-s_{{+-}}\equiv s_+$, where
\begin{equation*}
  s_-\equiv s_{+-} = \left(\frac{\alpha}{2\beta}\right)^{1/2}\left[1-\left(1+4\frac{4\beta}{\alpha^2}\varepsilon^* z\right)^{1/2}\right]^{1/2}\sim \ii\left(\frac{\varepsilon^* z}{\alpha}\right)^{1/2} \left[1-\frac{\beta}{2\alpha^{2}}\varepsilon^* z\right],
\end{equation*}
specifically in the scenario $z>0$. This is the case where the saddles mapping is complex, and arises for the same reason as for the turning point at $x=a$. We further recall that the notation for the saddles $s_\pm$ here is slightly different to the $x=a$ case. Specifically, the lower index is chosen such that $\Re\psi(s_+)<\Re\psi(s_-)$; choosing $z<0$ and performing the same analysis of the coalescing saddles yields the same approximation.

Since
\begin{equation*}
    -\psi(s_+)=\psi(s_-)\sim\frac{(\varepsilon^* z)^{3/2}}{\alpha^{1/2}}\left(\frac{2}{3}-\frac{\beta}{5}\frac{\varepsilon^* z}{\alpha^2}\right)>0,
\end{equation*}
we can conclude that $s_+\mapsto t_+$ as in the $x=a$ turning point.
Next, we note that
\begin{equation*}
  \psi''(s_+) \sim -2 (\alpha\varepsilon^* z)^{1/2} \left(1+ \frac{3}{2}\frac{\beta}{\alpha^2}\varepsilon^* z\right),
\end{equation*}
and
\begin{equation*}
  e^{i \sigma} |\Sigma| = \psi(s_-)-\psi(s_+) = 2 \psi(s_-) \sim 2\frac{(\varepsilon^* z)^{3/2}}{\alpha^{1/2}}\left(\frac{2}{3}-\frac{\beta}{5}\frac{\varepsilon^* z}{\alpha^2}\right).
\end{equation*}
Therefore
\begin{align*}
\sigma = 0, \qquad
|\Sigma|\sim \frac{4}{3}\frac{(\varepsilon^* z)^{3/2}}{\alpha^{1/2}} -\frac{2}{5} \frac{\beta}{\alpha^{5/2}} (\varepsilon^* z)^{5/2}, \qquad
\tau=0.
\end{align*}
which again rules out the second possibility ($\tau = 2\pi/3$), as above, %(avoiding Stokes line)
and
\begin{equation*}
  |\eta| = \left(\frac{3}{4} |\Sigma|\right)^{2/3} \sim \frac{\varepsilon^* z}{\alpha^{1/3}}\left(1- \frac{\beta}{5 \alpha^2} \varepsilon^* z\right),
\end{equation*}
where $\eta = |\eta|$, $t_\pm=\pm|\eta|^{{1/2}}$ and $A=0$. \bl{
Further, we note that
\begin{equation*}
    f(t_\pm) = \left(\frac{2 t_\pm}{\psi''(s_\pm)}\right)^{1/2} = \left(\frac{2 t_+}{\psi''(s_+)}\right)^{1/2} \sim  \dfrac{i}{\alpha^{1/3}} \left(1-\frac{4}{5}\frac{\beta}{\alpha^2} \varepsilon^* z\right),
\end{equation*}
and therefore}
\begin{align*}
   \Im(A_0)&=f(t_+)  \sim \alpha^{-1/3} \left(1-\frac{4}{5}\frac{\beta}{\alpha^2} \varepsilon^* z\right),\quad \Re(A_0)=0,
\end{align*}
and so the approximation across the turning point reads
\begin{equation} \label{eq.TurningPointSolu_x=b}
  \Re{p_-}(x) \sim 2 \pi (\varepsilon^*)^{1/3}A_0 \mathrm{Ai}\left(\eta (\varepsilon^*)^{-2/3}\right) \sim 2 \pi  (\varepsilon^*)^{1/2} \alpha^{-1/3} \mathrm{Ai}\left(\left(\frac{\rho}{\alpha}\right)^{1/3}\frac{x-b}{\varepsilon^{2/3}}\right).
\end{equation}

\subsection{Numerical evaluation of the contour integral solution along the SDC}\label{sec_numerical}

Taking advantage of explicit knowledge of the contour parameterisation for both $z>0$, $z<0$, we can numerically integrate the contour integral form of the solution in {\bl $z$} for the values of particular parameters. We compare these contour integral results for a range of values of {\bl $z$} with the analytic approximation of the contour integral both near the turning point (the coalescing saddles case) and far from the turning point.

\begin{figure}
\centering
 \begin{subfigure}{.495\textwidth}
   \centering
  \includegraphics[width=.95\linewidth]{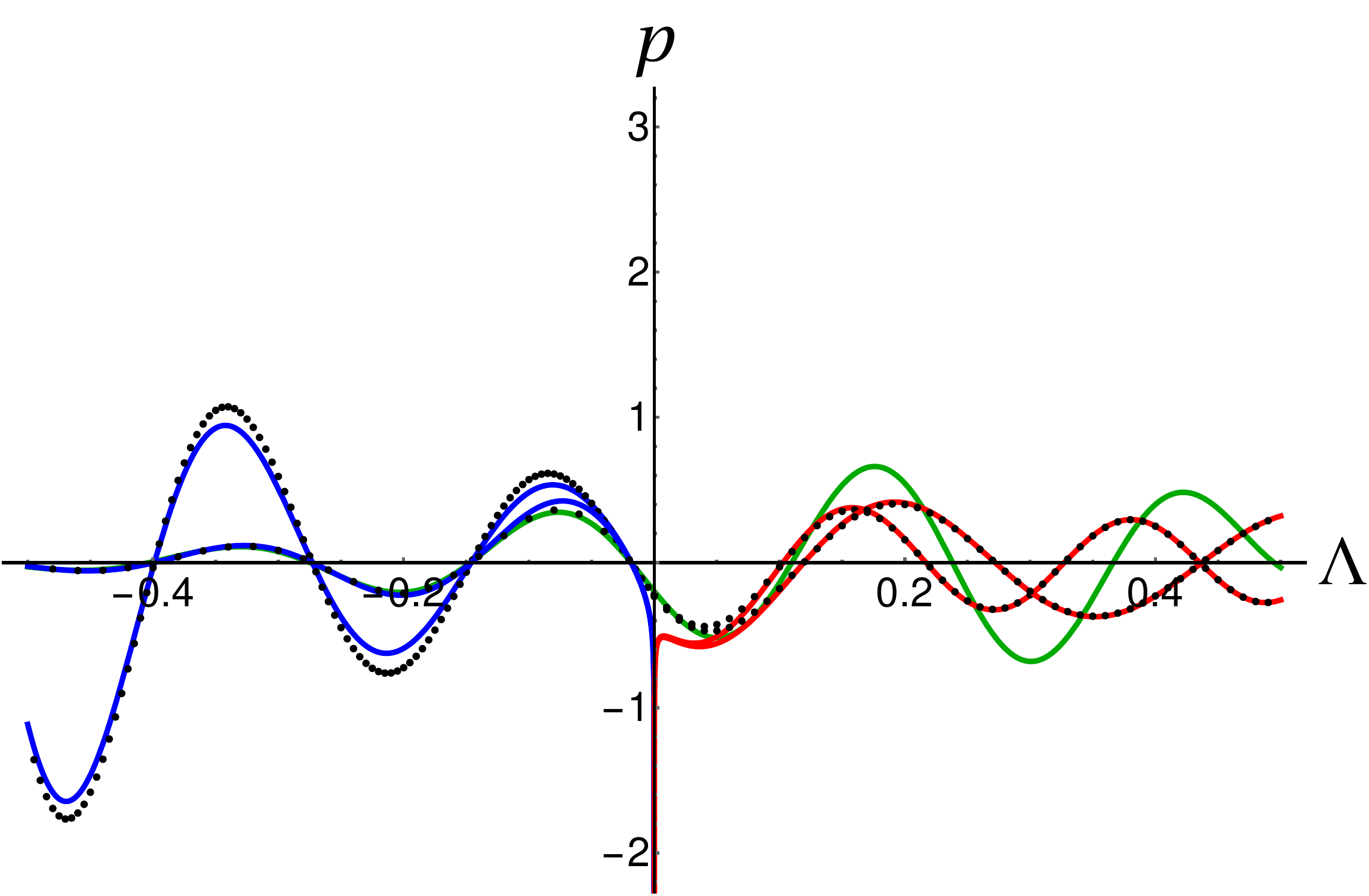}  
\caption{Parameter values taken as $\alpha =3$,  $\beta=1$, $\varepsilon = 5 \times 10^{-2}$.}%tady to pisu blbe v ozn souboru; eps 1p20 je zlomek 1/20
\end{subfigure}
\begin{subfigure}{.495\textwidth}
   \centering
  \includegraphics[width=.95\linewidth]{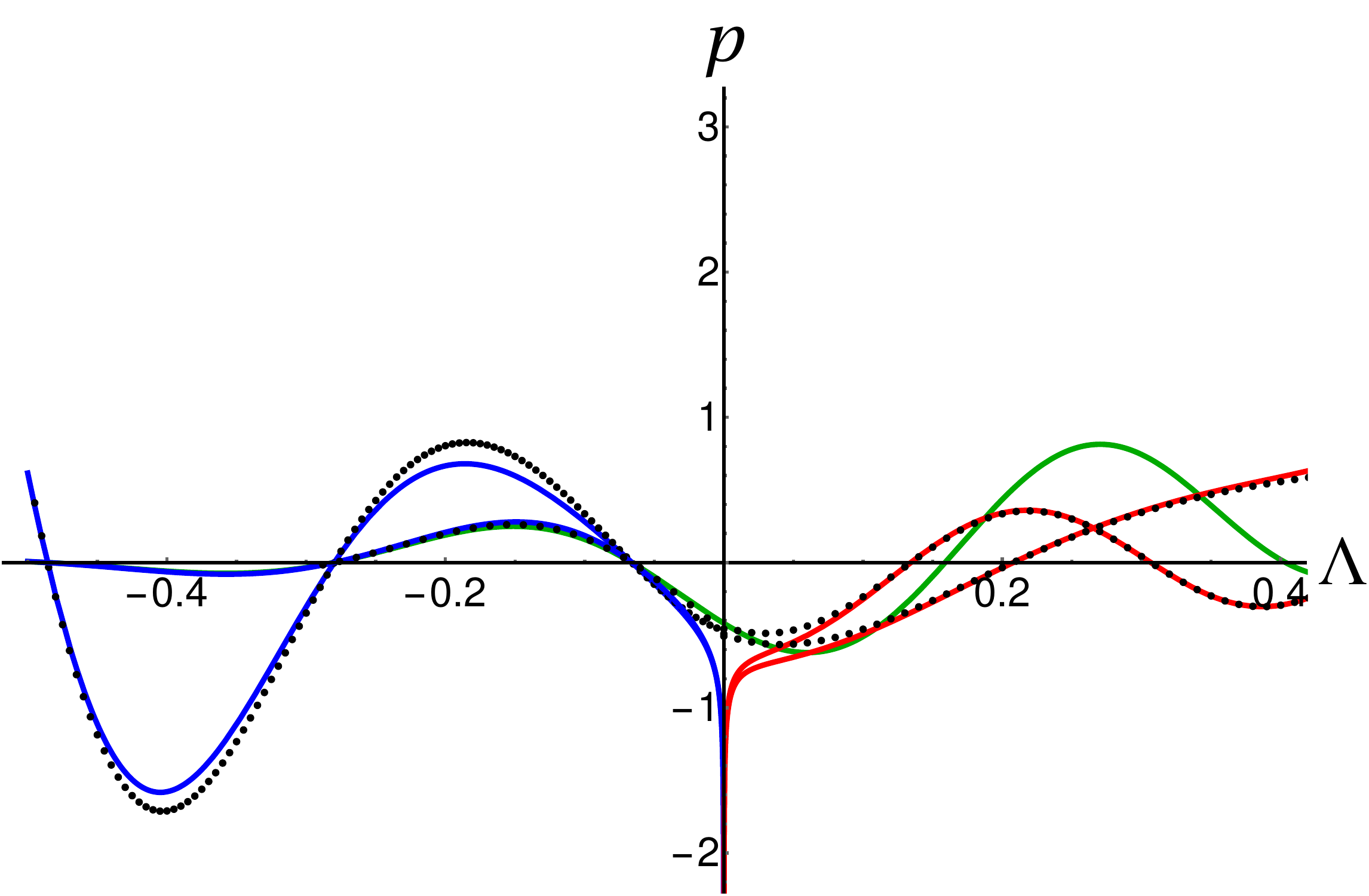}  
\caption{Parameter values taken as $\alpha =2$,  $\beta=2$, $\varepsilon =  5 \times 10^{-2}$.}
\end{subfigure}

\begin{subfigure}{.495\textwidth}
   \centering
  \includegraphics[width=.95\linewidth]{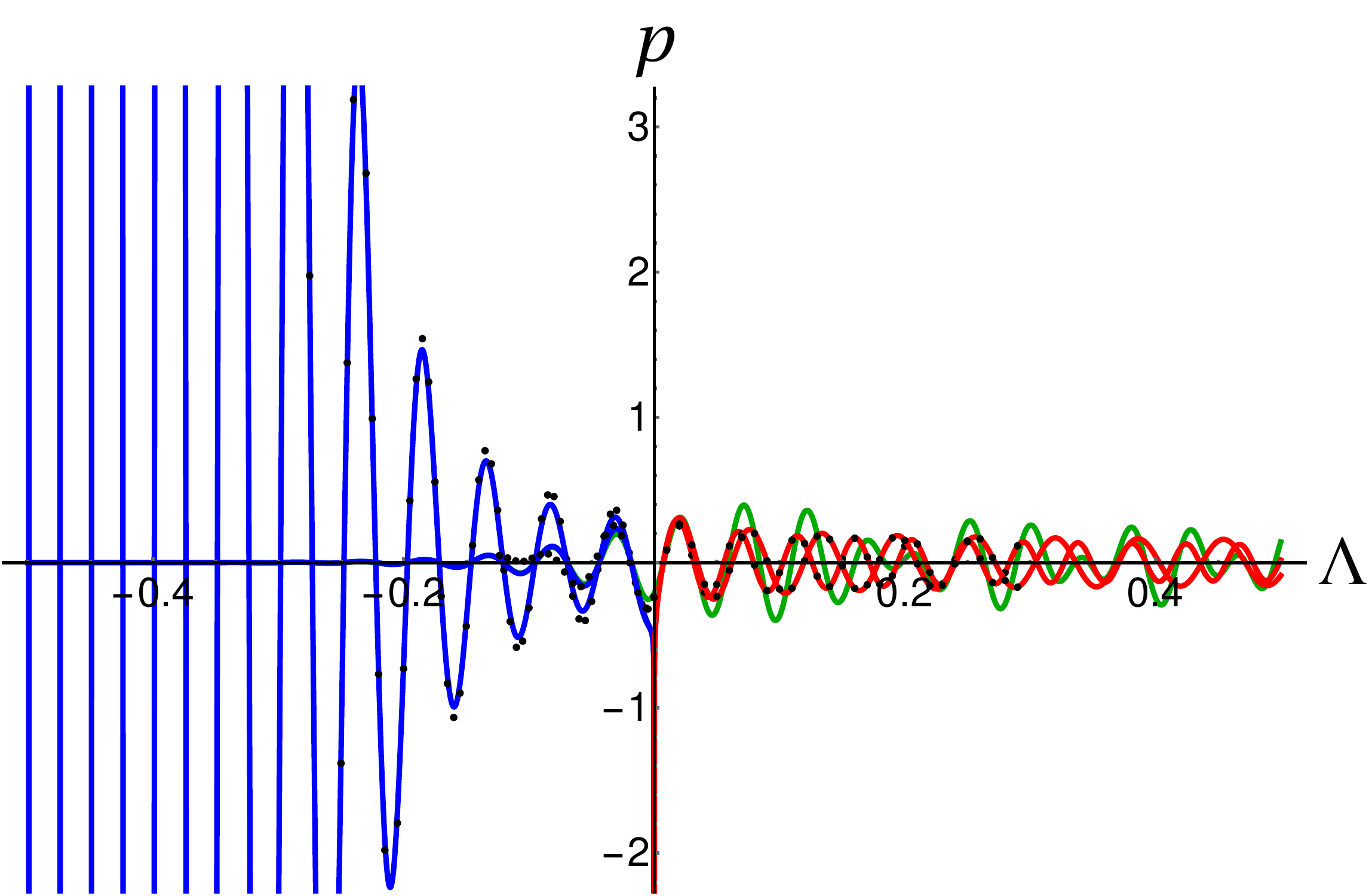}  
\caption{Parameter values taken as $\alpha =3$,  $\beta=1$, $\varepsilon = 10^{-2}$.}
\end{subfigure}
\begin{subfigure}{.495\textwidth}
  \centering
    \includegraphics[width=.95\linewidth]{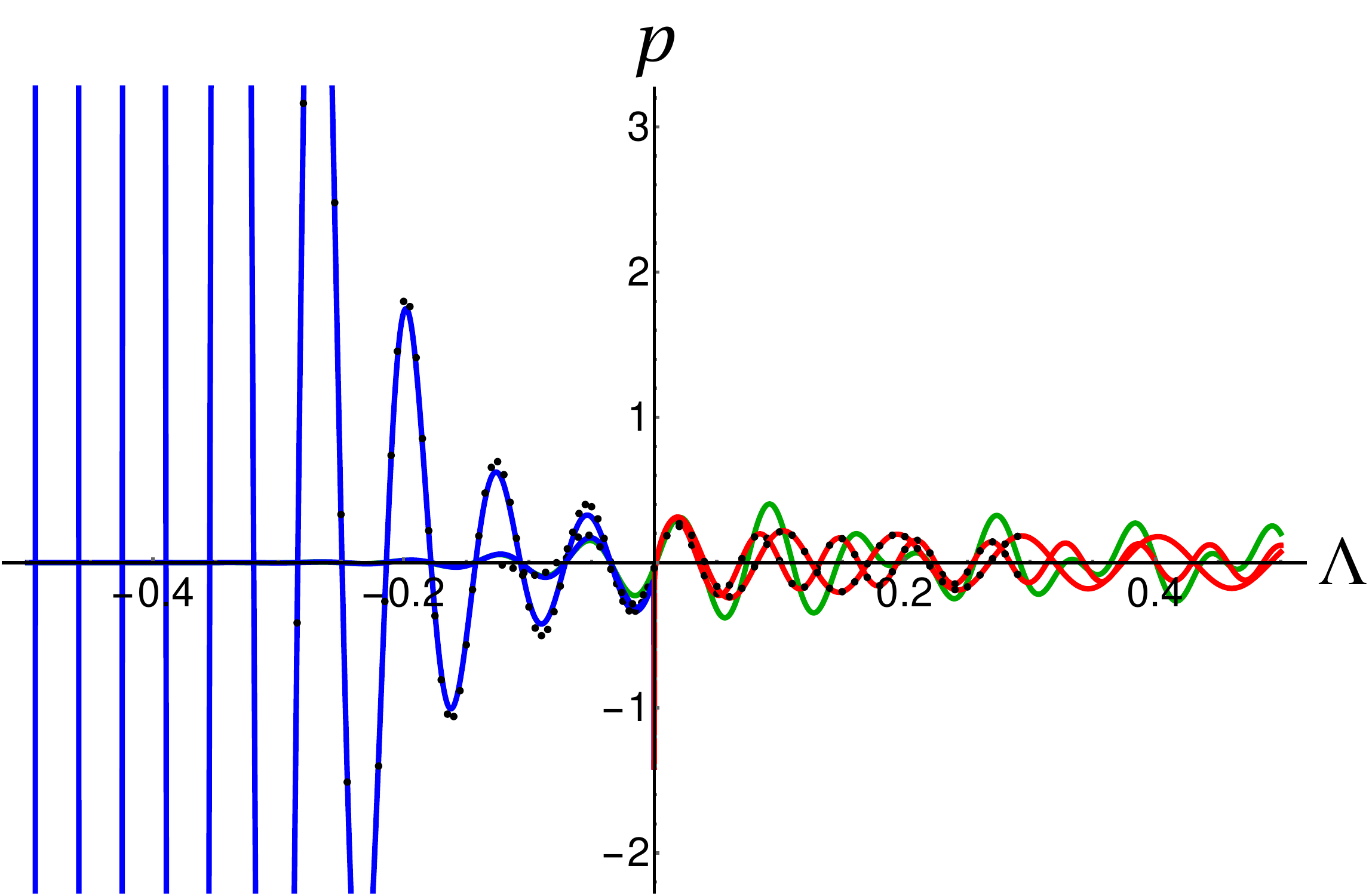}  
\caption{Parameter values taken as $\alpha =3$,  $\beta=2$, $\varepsilon = 10^{-2}$.}
\end{subfigure}
\caption{\label{fig:Verifying_xa} For the proximity of the turning point $x=a$, {\rd results for the cosine variant of the solutions arising from numerically integrating along} the  SDC contour are shown as dots and the analytical estimates of the contour integral outside of the turning point are shown as solid curves with the horizontal axis for the plots is in terms of $\Lambda=\rho\varepsilon z$.
	{\rd The cosine analytical solution} excluding the immediate vicinity of the turning point (solid line) transitioning from an exponentially decaying solution to oscillatory is a combination of $p_{+}$ branch for $z<0$ (in blue), and $p_{-}$ branch for $z>0$ (in red). In green is the approximation for the connection between these solutions across the turning point, via Eq.~\eqref{eq.TurningPointSolu_x=a_cos}.   Each dot is the numerically calculated contour integral, Eq.~\eqref{eq.innerp} along the SDC {$C_s$} for a given $\varepsilon^* z$.  
	  In addition, 
	the analytical solution {excluding the immediate vicinity of the turning point} (solid line) transitioning from an exponentially growing solution to oscillatory is a combination of $p_{-}$ branch for $z<0$ (in blue) and $p_{+}$ branch for $z>0$ (in red) is also shown, with dots giving the associated numerically calculated contour integral.
 Note that the analytical approximation developed without the use of coalescing saddles estimates the solution well once away from the turning point, while it diverges at the turning point, {\bl while the final coalescing saddle approximation captures the transition} from an exponentially growing to an exponentially decreasing solution, Eq.~\eqref{eq.TurningPointSolu_x=a}.}
%{Turning point $x=a$, results of the numerical  integration along the SDC contour shown as dots and the analytical estimates of the contour integral outside of the turning point as solid curves. Solid lines represent the approximate solutions: red for $z>0$ in Eq.~(\ref{eq.inner.x=a.L>0.solu}), blue for $z<0$ in Eq.~(\ref{eq.inner.x=a.L<0.solu}), green for the approximation across the turning point in Eq.~\eqref{eq.TurningPointSolu_x=a}. Each dot is the numerically calculated contour integral, Eq.~\eqref{eq.innerp} along the SDC {$C_s$} for a given $\varepsilon^* z$.  The analytical solution {excluding the immediate vicinity of the turning point} (solid line) transitioning from an exponentially decaying solution to oscillatory is a combination of $p_{+}$ branch for $z<0$ (in blue), and $p_{-}$ branch for $z>0$ (in red). The analytical solution {excluding the immediate vicinity of the turning point} (solid line) transitioning from an exponentially growing solution to oscillatory is a combination of $p_{-}$ branch for $z<0$ (in blue) and $p_{++}$ branch for $z>0$ (in red). Note that the approximation away from the turning point estimates the solution well, while it diverges at the turning point but where the final approximation captures the transition from an exponentially growing to an exponentially decreasing solution, Eq.~\eqref{eq.TurningPointSolu_x=a}.}
\end{figure}

For example, considering $x=a$ and $z>0$, with a contour parameterized by $r$, we have %{\mg VK: I think we dropped the definition of $\alpha$ somewhere around here: $\alpha$ is the value of $\Im\psi$ at the chosen saddle.}
\begin{align*}
  C_s\equiv \Bigg\{(a,b)\, \Bigg| \, a&=r,~~
  b=\pm \left(-1+r^2\pm\frac{1}{r}\left(\frac{4}{5} r^6 - \frac{4}{3} r^4 + \varepsilon^* z r^2 +\alpha r\right)^{1/2}\right)^{1/2}  \Bigg\}.
\end{align*}
\bl{First, we numerically evaluate the contour integral expression near the turning point $x=a$, that is the Eq.~\eqref{eq.innerp}. The obtained numerical values {\rd for the cosine solutions} are shown as dots, and we overlap those with the {\rd associated} smooth curves corresponding to the inner solutions and the resolved behaviour across the turning point using the method of coalescing saddles. We see that the analytical results nicely match the numerics for various parameter values in the respective range of {\rd $\Lambda=\varepsilon \rho z>0$} in Fig.~\ref{fig:Verifying_xa}. In particular, the inner solution (Eq.~\eqref{eq.inner.x=a.L>0.solu} for $z>0$, Eq.~\eqref{eq.inner.x=a.L<0.solu} for $z<0$) accurately  approximates the solution when sufficiently away from the turning point while the coalescing saddles method {\rd (Eq.~\eqref{eq.TurningPointSolu_x=a_cos} across the turning point)} performs well {\rd in the immediate vicinity of the} turning point but overestimates the amplitude modulation when away from the turning point $x=a$.
}

\bl{
We proceed similarly for the second turning point, $x=b$. In this case, we showed that we have a well-approximated solution already across the turning point already from the inner expression (corresponding to the solution $p_+$) which is purely oscillatory, Eq.~(\ref{eq.inner.x=b.L<0.solu}a). The contour passing through the $s_-$ saddle generates inner solution which is valid only when sufficiently away from the turning point $x=b$ if we take the $\cos$ variant of the amplitude oscillations; see Eq.~(\ref{eq.inner.x=b.L<0.solu}a) and (\ref{eq.inner.x=b.L>0.solu}) and the nearby discussion. Hence, only {\rd for this solution has   it been} necessary to resolve the behaviour across the turning point using the coalescing saddles method, which resulted in the expression Eq.~\eqref{eq.TurningPointSolu_x=b}. In this case, the approximate solution is the Airy function without any modulation and with a different scaling of the argument ($2^{1/3}$ larger) than across the $x=a$ turning point.
}

\begin{figure}
\centering
\begin{subfigure}{.495\textwidth}
   \centering
  \includegraphics[width=.95\linewidth]{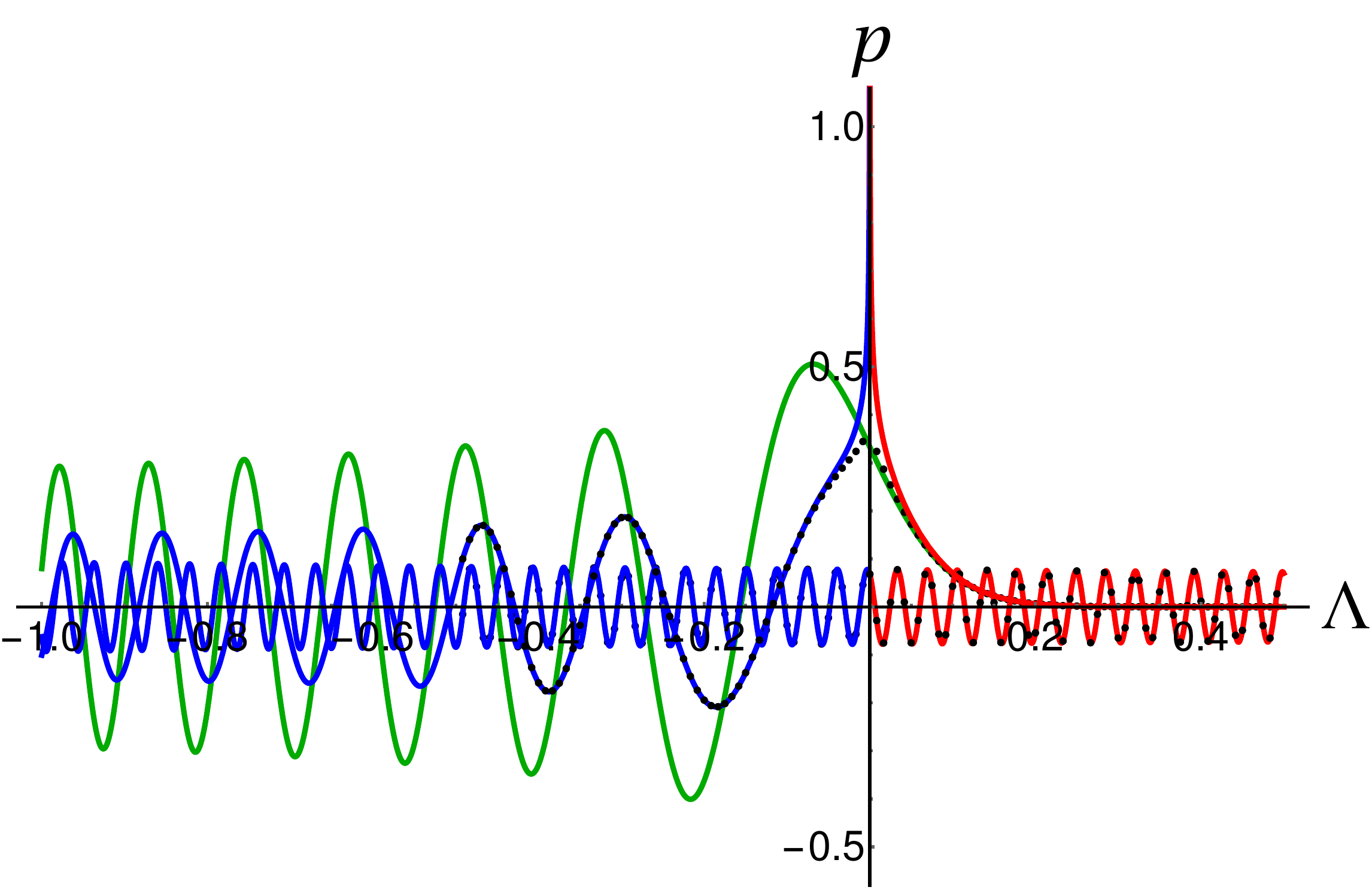}  
\caption{Parameter values taken as $\alpha =3$,  $\beta=1$, $\varepsilon = 10^{-2}$.}
\end{subfigure}
\begin{subfigure}{.495\textwidth}
  \centering
    \includegraphics[width=.95\linewidth]{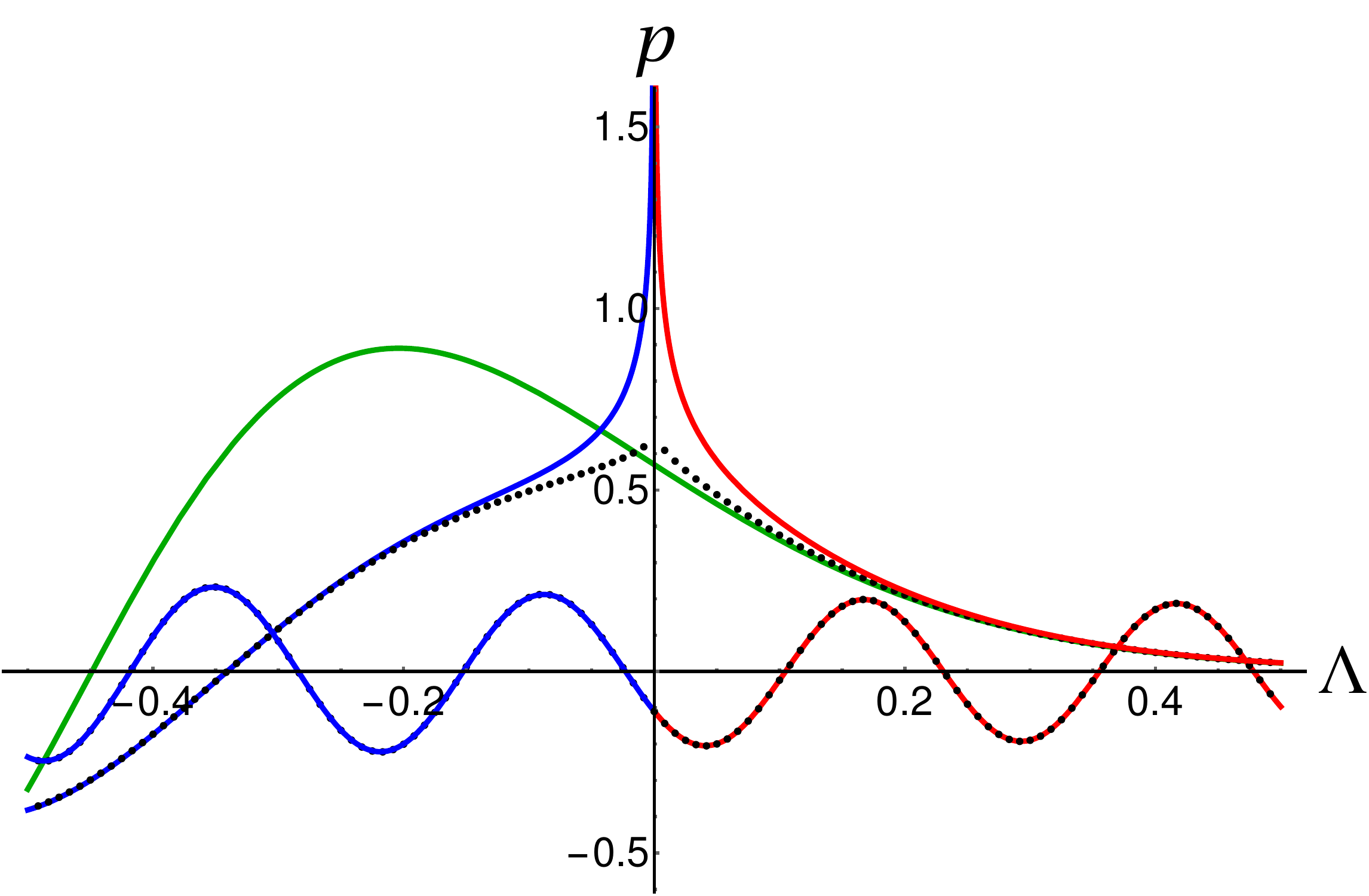}  
\caption{Parameter values taken as $\alpha =3$,  $\beta=2$, $\varepsilon = 5\times 10^{-2}$.}
\end{subfigure}
\caption{\label{fig:Verifying_xb}
{\rd   For the proximity of the turning point $x=b$, results for the cosine variant of the solutions arising from numerically integrating along the  SDC contour  are shown as dots    in terms of $\Lambda=\rho\varepsilon z$. The solid lines represent the associated approximate solutions: red for $z>0$ in Eq.~(\ref{eq.inner.x=b.L>0.solu}), blue for $z<0$ in Eq.~(\ref{eq.inner.x=b.L<0.solu}).  In contrast to the turning point at $x=a$ we have a uniformly accurate solution across the turning point already from the inner expression, which corresponds to the $p_+$ solution and is oscillatory, Eq.~(\ref{eq.inner.x=b.L<0.solu}a). However, the contour passing through the $s_-$ saddle, and thus the $p_-$ solution,  generates an inner solution that is uniformly valid only for the sine variant; for the cosine variant accuracy is only ensured 
sufficiently far from the turning point $x=b$, see Eq.~(\ref{eq.inner.x=b.L<0.solu}b) and (\ref{eq.inner.x=b.L>0.solu}) and the discussion there. Hence, only for this solution has it been  necessary to resolve the behaviour across the turning point using the coalescing saddles method.  This final aspect of the approximation in this case is depicted  by the green solid curve, as derived from Eq.~\eqref{eq.TurningPointSolu_x=b} and captures the transition from an exponentially growing to an exponentially decreasing solution. It is given by an Airy function without any modulation, but with a larger scaling of the argument compared to the analogous solution across  the $x=a$ turning point.}}
\end{figure}

%%%%%%%%%%%%%5
\section{Multiple-scales approach for resolving the inner solution} \label{sec.multiple_scales}

An alternative way to obtain the behaviour across turning points is to follow the methodology of \cite{dalwadi2021emergent,dalwadi2023universal} for related inner problems, and use a multiple-scales-type approach. This requires introducing two inner spatial scales near the turning point, a fast scale $x-a = O(\varepsilon)$ which captures the oscillation, and an intermediate scale $x-a = O(\varepsilon^{2/3})$ which captures the amplitude variation. The reason the latter scale is `intermediate' is because it falls between the fast scale and the `slow' outer scale where $x-a = O(1)$. We note that this approach is only feasible near the turning point here (i.e.~not within the outer regions) because the frequency is constant to leading-order within the inner region.

The inner equation \eqref{eq.inner} is valid near $x=a$, and is given in terms of the fast scale $y = (x-a)/\varepsilon$. Following the scalings in \cite{dalwadi2023universal}, we introduce the additional intermediate scaling
\begin{align}
\sigma = \varepsilon^{1/3} y,
\end{align}
noting that this is equivalent to $\sigma = (x-a)/\varepsilon^{2/3}$. Proceeding via the method of multiple scales \cite{Bender1999}, we consider $p = p(y,\sigma)$, noting that the use of $\sigma$ here is unrelated to its use in the previous section. The additional degree of freedom this introduces will be removed later by imposing appropriate periodicity constraints in $y$. Under the standard method of multiple scales assumption, we treat $y$ and $\sigma$ as independent, which transforms the derivative as
\begin{align}
\label{eq: MoMS time deriv}
\dfrac{\mathrm{d}}{\mathrm{d}y} \mapsto \dfrac{\partial}{\partial y} + \varepsilon^{1/3} \dfrac{\partial}{\partial \sigma}.
\end{align}
Using the transformation \eqref{eq: MoMS time deriv} in \eqref{eq.inner}, we obtain
\begin{subequations}
\label{eq.inner transform}
\begin{align}
  0 = L[p] + \varepsilon^{2/3} \sigma \rho p - 2 \varepsilon^{1/3} \dfrac{\partial^2}{\partial \sigma \partial y} \left(\alpha + 2 \beta \dfrac{\partial^2}{\partial y^2} \right)p - \varepsilon^{2/3} \dfrac{\partial^2}{\partial \sigma^2} \left(\alpha + 6 \beta \dfrac{\partial^2}{\partial y^2} \right)p,
\end{align}
where
\begin{align}
  L[p] = \left(1-\frac{\alpha^2}{4\beta}\right)p - \left(1+\alpha \frac{\partial^2}{\partial y^2}+\beta \frac{\partial^4}{\partial y^4}\right)p.
\end{align}
\end{subequations}
In \eqref{eq.inner transform}, we have written $\varepsilon y = \varepsilon^{2/3} \sigma$ to be consistent with the implicit requirement underlying the method of multiple scales i.e.~that the solution is periodic in $y$, and any variation away from this periodicity occurs via the $\sigma$ variable.

Next we expand $p$ as a power series in $\varepsilon^{1/3}$:
\begin{align}
p(y,\sigma) \sim p_0(y,\sigma) + \varepsilon^{1/3} p_1(y,\sigma) + \varepsilon^{2/3} p_2(y,\sigma),
\end{align}
substitute this into \eqref{eq.inner transform} and equate coefficients of powers of $\varepsilon^{1/3}$. At leading order, we obtain
\begin{align}
\label{eq: LO eq for MoMS}
L[p_0] = 0,
\end{align}
with accompanying periodicity constraints in $y$. Eq.~\eqref{eq: LO eq for MoMS} is a linear differential equation with constant coefficients, and so it is straightforward to obtain the solution
\begin{align}
\label{eq: p_0 sol}
p_0(y,\sigma) = A(\sigma) e^{i k y} + \text{c.c.},
\end{align}
with
\begin{align}
\label{eq: k def}
k = \sqrt{\alpha/2\beta},
\end{align}
noting that the remaining two linearly independent solutions of \eqref{eq: LO eq for MoMS} have been neglected in \eqref{eq: p_0 sol} since they are not periodic in $y$. Additionally, we note that $A(\sigma)$ is the (currently undetermined) slowly varying amplitude of the leading-order solution and $\text{c.c.}$ denotes the complex conjugate. Our remaining task is to determine the governing equation for $A(\sigma)$. To do this, we must proceed to higher asymptotic orders and derive an appropriate solvability condition.

At next order ($O(\varepsilon^{1/3})$) in \eqref{eq.inner transform}, we obtain the system
\begin{align}
\label{eq: first correction eq for MoMS}
L[p_1] = 2 \dfrac{\partial^2}{\partial \sigma \partial y} \left(\alpha + 2\beta \dfrac{\partial^2}{\partial y^2} \right) p_0 = 0,
\end{align}
where the right-hand side vanishes immediately upon substituting in the leading-order solution \eqref{eq: p_0 sol}. Therefore, we may solve \eqref{eq: first correction eq for MoMS} in the same way as the leading-order solution to obtain
\begin{align}
\label{eq: p_1 sol}
p_1(y,\sigma) = B(\sigma) e^{i k y} + \text{c.c.},
\end{align}
where $B(\sigma)$ is another undetermined function of the intermediate variable. However, in contrast to $A$, we will not require any knowledge of $B$ to obtain the results we seek, so we do not determine it.

The solvability condition we are seeking arises at the subsequent order ($O(\varepsilon^{2/3})$) in \eqref{eq.inner transform}, from which we obtain the system
\begin{align*}
%\label{eq: first correction eq for MoMS}
L[p_2] &= -\sigma \rho p_0 + 2 \dfrac{\partial^2}{\partial \sigma \partial y} \left(\alpha + 2\beta \dfrac{\partial^2}{\partial y^2} \right) p_1 + \dfrac{\partial^2}{\partial \sigma^2} \left(\alpha + 6 \beta \dfrac{\partial^2}{\partial y^2} \right)p_0, \notag \\
&= -\sigma \rho A e^{i k y} - 2 \alpha \dfrac{\partial^2 A}{\partial \sigma^2} e^{i k y} + \text{c.c.}
\end{align*}
The solvability condition here arises from requiring $p_2$ to be periodic in $y$, which requires the coefficients of $e^{iky}$ to vanish. Namely, this requirement yields the following governing equation for $A$:
\begin{align}
\label{eq: amplitude equation}
2 \alpha \dfrac{\partial^2 A}{\partial \sigma^2} + \sigma \rho A = 0,
\end{align}
which, with accompanying matching conditions, is a closed system for $A$. We note that \eqref{eq: amplitude equation} is a sublimit of the generalized (modified) normal form for spatio-temporal heterogeneities across pitchfork and transcritical bifurcations derived and analysed in \cite{dalwadi2023universal}.

With the constraint that the outer solution is not exponentially large as $\sigma \to -\infty$ (which can be formally justified via asymptotic matching), the general solution to \eqref{eq: amplitude equation} is given in terms of the Airy function:
\begin{align}
\label{eq: amplitude solution}
A(\sigma) = \gamma \mathrm{Ai}\left( - \left(\dfrac{\rho}{2 \alpha}\right)^{1/3} \sigma \right),
\end{align}
where $\gamma$ is a (complex) constant determined via matching. Substituting \eqref{eq: amplitude solution} into the leading-order representation \eqref{eq: p_0 sol} and writing it in terms of the outer variable $x$, we obtain the leading-order inner solution
\begin{align}
\label{eq: p_0 sol with amplitude}
p_0(y,\sigma) = \gamma \mathrm{Ai}\left( - \left(\dfrac{\rho}{2 \alpha}\right)^{1/3} \dfrac{x - a}{\varepsilon^{2/3}} \right) \exp \left(\dfrac{i (x-a)}{\varepsilon} \sqrt{\dfrac{\alpha}{2 \beta}} \right) + \text{c.c.}.
\end{align}
{\bl In particular, the form of this solution is precisely that of Eq.~\eqref{eq.TurningPointSolu_x=a}, and 
for real $\gamma$, of the cosine type solution given by  Eq.~\eqref{eq.TurningPointSolu_x=a_cos}, demonstrating that the multiple scales method generates the same solution as the coalescing saddle method. With either solution, we finally note that} to obtain appropriate matching conditions, we write $\gamma = R \exp(i \Theta)$ and use the standard large-argument asymptotic representations of the Airy function to obtain the following far-field representations of the inner solution \eqref{eq: p_0 sol with amplitude}:
\begin{align}
p &\sim \left( \dfrac{2 \alpha}{\rho} \right)^{1/12} \dfrac{R \varepsilon^{1/6}}{\pi^{1/2} (a-x)^{1/4}} \exp\left[-\dfrac{2}{3} \left(\dfrac{\rho}{2 \alpha} \right)^{1/2} \dfrac{(a-x)^{3/2}}{\varepsilon} \right] \cos \left(\left( \dfrac{2 \alpha}{\rho} \right)^{1/2} \dfrac{a-x}{\varepsilon} - \Theta \right) \quad \text{as } x \to a^{-}, \\
p &\sim \left( \dfrac{2 \alpha}{\rho} \right)^{1/12} \dfrac{2 R \varepsilon^{1/6}}{\pi^{1/2} (x-a)^{1/4}} \sin\left[\dfrac{2}{3} \left(\dfrac{\rho}{2 \alpha} \right)^{1/2} \dfrac{(x-a)^{3/2}}{\varepsilon} + \dfrac{\pi}{4} \right] \cos \left(\left( \dfrac{2 \alpha}{\rho} \right)^{1/2}  \dfrac{x-a}{\varepsilon} + \Theta \right) \quad \text{as } x \to a^{+}.
\end{align}
These conditions could be used to match with the outer solution \eqref{eq.p_out}, and obtain the appropriate connection formulae for the outer solutions across the turning point.

%\todo{projit todosy; explicit comparison of the two approaches; zkontrolovat AJ pomoci writefullu; navrhnout discussion; AI abstract - poslat vsem s tim, ze zadam Andrewa o prvni cteni a upravy zejm pak Intro, Discu; precist si Moa!}

%%%%%%%%%%%%%%%%%%%%%%%%%%% discussion
\section{Discussion}\label{discussion}

The WKBJ outer solution has been demonstrated to be a useful tool in studying linear pattern-forming models forced via spatial heterogeneity \cite{krause2020, gaffney2022spatial, krause2024pattern}. In such settings, it was demonstrated that for the system to exhibit a Turing-like instability, the second turning point, $x=b$, must necessarily be ruled out. However, the turning point $x=a$ is found to be indispensable for the selection of an unstable mode and its support, which changes with the wavenumber. It is noteworthy that the $x=a$ turning point invariably corresponds to the lower value of $r(x)$ in comparison to $x=b$ and thus represents the first point of transition when instability manifests, and hence the location of a bifurcation-point from a quasi-static view, as explored in \cite{krause2024pattern}. Nevertheless, the knowledge of the transition across these turning points is essential in order to construct an approximate global solution.

In this study, the focus was on calculating the asymptotic solution of the spatially varying Swift-Hohenberg problem, including the behaviour of solutions near both types of turning points. To this end, an integral form of solution was considered, and subsequently approximated. To capture the behaviour in the neighbourhood of turning points, the method of coalescing saddles was employed. The methodology employed in this study is comprehensive, including the determination of complex branch selection of roots, a direct consequence of the local mapping to the cubic polynomial considered by Chester et al \cite{chester1957extension}. The integration of these results has enabled us to successfully approximate the global solution to the heterogeneous Swift-Hohenberg problem. This approximation was confirmed by comparison to numerical integration of the contour integral form of the solution.

Finally, we employed the recent insights of \cite{dalwadi2023universal}, where the modified normal forms for spatio-temporal variations across transcritical and supercritical pitchfork bifurcations were derived and analysed using a multiple-scales analysis. The turning point at $x=a$ here is a sublimit of the modified normal forms in \cite{dalwadi2023universal}. Hence,  through careful choice of the intermediate scaling here, we were able to use the multiple scales approach of \cite{dalwadi2023universal} to circumvent the technically intricate method of coalescing saddles for the $x=a$ turning point. This approach generates the same leading-order behaviour across the turning point as the method of coalescing saddles, but in a more concise manner, {\bl leading to the hypothesis that the method of coalescing saddles might be analogously circumvented more generally, even if only for certain classes of integral asymptotics.}
However, since the local frequencies of the outer solutions become degenerate near the $x=b$ turning points, further investigation would be required to consider adapting this approach near $x=b$.

The approach presented here, and in all other uses of WKBJ asymptotics for pattern-forming models that we are aware of, is restricted to one spatial dimension. There are fundamental difficulties even constructing outer solutions in heterogeneous multi-dimensional problems, although numerically they can often behave as one would expect from a naive interpretation of the one-dimensional theory \cite{woolley2021bespoke}.  
There are at least two other direct extensions of the work presented here which would be especially interesting to pursue. Firstly would be to explore the possibility that the multiple-scales approach of Section \ref{sec.multiple_scales} can in general be applied to integrals of the form \eqref{eq.innerp} by first writing the associated differential equations that they solve. In this way, one could in principle subsume coalescing saddles arguments, and possibly Laplace-like approaches to integral asymptotics more generally. Secondly, in \cite{krause2024pattern} the authors numerically showed a breakdown of the linear theory for nonlinearities $N(u)$ which `locally' bifurcated according to a subcritical Turing instability. One could consider a quintic generalisation of the weakly nonlinear theory presented in \cite{dalwadi2023universal} as a possible method for capturing such nonlinear effects in the spatially-forced context.

%%%%%%%%%%%%%%%%%%%%%%%% appendix %%%%%%%%%%%%%%%%%%%%%%%%%%%%
\newpage
\appendix

\bibliographystyle{plain} 
\bibliography{refs} 

\end{document}